\documentclass[aps,twocolumn,floats,prd,nofootinbib,superscriptaddress,showkeys,10pt]{revtex4-2}

\usepackage{amsfonts,amsmath,amssymb} 
\usepackage{graphicx}

\newcommand{\sgn}{\text{sgn}}

\usepackage{hyperref}
\hypersetup{
	colorlinks=true,
	allcolors=blue,
	linktocpage=true,
	pdfhighlight=/N
}

\begin{document}
	
\title{Spherically symmetric configurations in the quadratic $f(R)$ gravity
}
\author{V.~I.~\surname{Zhdanov}}
\email{valery.zhdanov@knu.ua}
\affiliation{Taras Shevchenko National University of Kyiv,  Kyiv 01601, Ukraine}
\author{O.~S.~\surname{Stashko}}\email{alexander.stashko@gmail.com}
\affiliation{ Department of Physics, Princeton University, Princeton, USA}
\affiliation{Goethe Universit\"{a}t, Max-von-Laue Str. 1, Frankfurt am Main, 60438, Germany}

\author{Yu.~V.~\surname{Shtanov}}
\email{shtanov@bitp.kyiv.ua}
\affiliation{Bogolyubov Institute for Theoretical Physics,  Metrologichna St.~14-b, Kyiv 03143, Ukraine} %
\affiliation{Taras Shevchenko National University of Kyiv,  Kyiv 01601, Ukraine} %

\begin{abstract}
We study spherically symmetric  configurations of the  quadratic $f(R)$ gravity ($f(R)=R-R^2/6\mu^2$). In case of a purely gravitational system, we  have fully investigated the global  qualitative behavior of all static solutions   satisfying the conditions of asymptotic flatness. These solutions are proved to be regular everywhere  except for a naked singularity at the center;  they are uniquely determined by the total mass $\mathfrak{M}$ and the ``scalar charge'' $Q$ characterizing the strength of the scalaron field at spatial infinity. The case $Q=0$ yields the Schwarzschild  solution, but an arbitrarily small $Q\ne 0$ leads to the appearance of a central naked singularity having a significant effect on the neighboring region, even when the space-time metric in the outer region is practically insensitive to the scalaron field. Approximation procedures are developed to derive asymptotic relations near the naked singularity and at spatial infinity, and the  leading terms of the solutions are presented.   
We have investigated the linear stability of the static solutions with respect to radial  perturbations satisfying the null Dirichlet boundary condition at the center and  numerically estimate the range of parameters corresponding to stable/unstable configurations. In particular, the configurations  with sufficiently small $Q$ turn out to be linearly unstable.
		
\end{abstract}


\maketitle{Keywords: modified gravity, $f(R)$ gravity, scalar fields, naked singularities}

\section{Introduction} \label{Introduction}

Modification of General Relativity (GR) by a Lagrangian in the form of a nonlinear function $f (R)$ of the scalar curvature $R$ is, perhaps, the simplest one and has long been the subject of numerous studies and applications  (see \cite{Sotiriou:2008rp, DeFelice:2010aj, Nojiri:2017ncd} for reviews). Compared to GR, such $f (R)$ gravity theory contains one extra scalar degree of freedom (the scalaron), which can be used for modelling a wide variety of phenomena, from the early inflationary regime \cite{Starobinsky:1980te, Vilenkin:1985md, Shtanov-Sahni2023, Odintsov-2023}, consistent with current observations \cite{Planck:2018jri}, to dark energy and dark matter at later epoch \cite{Capozziello:2006uv, Nojiri:2008nt, Cembranos:2008gj, Cembranos:2015svp, Corda:2011aa, Katsuragawa:2016yir, Katsuragawa:2017wge, Yadav:2018llv, Parbin:2020bpp, Shtanov:2021uif, Shtanov:2022xew}.

Natural questions arise about the possible effects of modified gravity in models of isolated systems that can simulate real astrophysical objects. Isolated configurations have received considerable attention in both GR and modified theories of gravity (see, e.g.,   \cite{Hawking1972,Bekenstein1972, Bekenstein1972a, Whitt:1984pd, 1995PhRvD..51.6608B,  Holdom2017, Kase2019,  Hernandez2020, Doneva_2020, Yamada-2020, Hong:2020miv, Bogush2022, Nguyen:2022xvh, Bakopoulos2022, DeFelice2023, DeFelice2023PLB,  Ye2023} and references therein).  {A number of  papers concern spherically symmetric (SS) configurations with either asymptotically flat or de~Sitter asymptotics.}

 {In this paper, we consider asymptotically flat spherically symmetric configurations in the well-known quadratic model\footnote{The metric signature is $(+\,, -\,, -\,, -)$, and the curvature conventions are $R^\alpha{}_{\beta\gamma\delta}= 
\partial_\gamma\Gamma^\alpha{}_{\beta\delta}- \ldots$, $R_{\mu\nu}=R^\alpha{}_{\mu\alpha\nu}$.} 
\begin{equation}\label{R2}
    f(R)=R-\frac{R^2}{6\mu^2}\, ,
\end{equation}
where $\mu$ is the scalaron mass. Correspondingly, we neglect the small cosmological constant.} 
This is one of the  best  $f(R)$ models  in the light of the present-day  observational data; it leads to a flattened scalaron potential realizing inflation in the early Universe. On the other hand, this model can be viewed as a first step to a wider class of theories  (see, e.g., \cite{Shtanov-Sahni2023}).

It is well known that the equations of the $f(R)$ gravity (Jordan frame) can be rewritten in  a scalar-tensor form  (Einstein frame) \cite{Sotiriou:2008rp, DeFelice:2010aj, Nojiri:2017ncd, Whitt:1984pd}.  { This is possible if $f(R)$ satisfies certain conditions\footnote{See also Appendix~\ref{generalizations}.} that are fulfilled in the case of (\ref{R2}).  We emphasize  that, in this paper, we consider the original  metric in the Jordan frame as a physical one, whereas the transition to the Einstein frame is a mathematical trick and the metric in the Einstein frame serves as  an auxiliary tool.}

Transition to the Einstein frame raises a number of questions inherent in scalar-tensor theories. One of them  concerns the no-hair theorems that prohibit   static black hole configurations  with a  non-trivial scalar field (SF) \cite{Hawking1972,Bekenstein1972, Bekenstein1972a, 1995PhRvD..51.6608B,Graham2014,Doneva_2020}.  
A closely related direction  deals with the ``no-go theorems,'' which prohibit   regular  asymptotically flat configurations for SF  \cite{Galtsov2001,Bronnikov2002}. 
These results involve  some   conditions concerning the regularity of solutions, the SF potentials and the coupling between SF and matter.  
Naturally, similar questions arise in the $f(R)$ gravity. Most attention has been paid to the existence of black holes (see, e.g., \cite{Cruz-Dombriz2009,Bhattacharya2016,Canate_2018,Nashed2020,Nashed2021}  and references therein). The no-hair theorems  of  
\cite{Bhattacharya2016, Canate_2018} strongly restrict possible black hole solutions in the $f(R)$ gravity.  {In particular, in the case of the quadratic $f(R)$ model \eqref{R2} with zero cosmological constant, the only spherically symmetric black hole configuration in vacuum is that described by the usual Schwarzschild solution\footnote{In case of a non-zero $\Lambda$ this is replaced by the Schwarzschild--de~Sitter solution}  \cite{Whitt:1984pd, Canate_2018}.} 

This does not mean that there are no SS configurations in this concrete model different from the solutions to GR; however, these are configurations with a naked singularity (NS) \cite{Hernandez2020} rather than a horizon. Systems with NS, if they really exist in nature, could be extremely interesting. Indeed, in the modern cosmological epoch, the effects of the modified gravity are expected to be typically very very weak in astrophysical objects, but they  can manifest themselves in the extreme conditions near NS. 
On the other hand, there is a widespread opinion that  systems with  NS  do not exist in the universe according to the  Penrose's  Cosmic Censorship  hypothesis \cite{Penrose1965,Penrose2002}. This hypothesis has never been proven, and its discussion shifts to issues of stability and fine-tuning of the input data  \cite{Christodoulou1984,Ori1987,Joshi1993,Joshi_2013, Ong2020}. 
Apparently, the existence and stability of NS depends on the types of  configurations involved and,  perhaps,  on the specific parameter regions in concrete models \cite{CLAYTON1998131, Gibbons2005,  Gleiser2006, Dotti_2022}.

Note that the analysis of static   solutions in $f(R)$ gravity in \cite{Hernandez2020}, where the existence of NS at the center was pointed out,  is essentially based on numerical computations. Some    assumptions of \cite{Hernandez2020} may be questioned or at least require rigorous justification. First of all, one must be sure that regular solutions with all possible configuration parameters satisfying the assumption of  asymptotic flatness do exist for all positive values of the radial variable $r$. Here it is necessary to pay attention to the exclusion of singularities for finite values of the radial variable; indeed, there are examples of static SS  configurations where ``spherical" singularities arise at finite values of $r>0$ \cite{ZhdSt, Strongly2021}.  Another problem that requires proper justification concerns the asymptotic properties of solutions, which is necessary to study the linear stability  considered in this work. The study of these issues is the subject of this article.
 
Our paper is organized as follows. In Section \ref{basic}, we review the general relations of $f(R)$ gravity as regards the transition to the Einstein frame. Section \ref{quadratic f(R) static} concentrates on the   {quadratic} $f (R)$ model. Field equations in the Einstein frame are written in Section \ref{SSS-Ein_frame} for the case of a non-stationary SS metric. 
Section \ref{sec:asympt} deals with the purely gravitational system; we prove that \textit{all} SS solutions with nonzero scalar field satisfying the asymptotic-flatness conditions, exist for all positive values of a radial variable and may have singularity only at the center. In this Section, asymptotic formulas near NS are obtained. In Section \ref{SS_perturbations} we write down equations for perturbations and discuss the boundary conditions. In Section \ref{num}  we use numerical calculations to illustrate the static solutions (see Figs. \ref{fig:metricJordanA} -- \ref{p3.png})  and to study their linear stability. In Section \ref{discussion}, we discuss our results. The details concerning the proof of  the existence and  uniqueness and justification of approximation methods are given in  appendices. 
 
 \section{Basic equations and notation}\label{basic}
 
In $f(R)$ gravity,  the standard Lagrangian of General Relativity in the  gravitational action  is replaced by a more general function\footnote{We use the units in which $c = \hbar = 1$ }   of the scalar curvature $R$:    
\begin{equation*} 
S_g=-\frac{1}{2\kappa} \int d^{4} x \, \sqrt{-g} \,f(R)   \, , 
\end{equation*}
where $\kappa=8\pi G$, and $G$ is Newton's gravitational constant.  
The corresponding dynamical equations   for the physical metric $g_{\mu\nu}$ (Jordan frame)  are \cite{Sotiriou:2008rp, DeFelice:2010aj, Nojiri:2017ncd} 
\begin{equation}\label{equations-f(R)}
f'(R)R_{\mu\nu}-\frac12 g_{\mu\nu}f(R)+\left( g_{\mu\nu} \Box -\nabla_\mu \nabla_\nu \right) f'(R) = \kappa T_{\mu\nu } \,   ,
\end{equation}
where  $T_{\mu\nu}=T_{\nu \mu}$ is the    energy-momentum tensor of non-gravitational fields   satisfying the covariant conservation law
\begin{equation}\label{energy-momentum_conservation}
\nabla _\nu T^{ \mu\nu}=0\,.
\end{equation}

There is a well-known procedure \cite{Sotiriou:2008rp, DeFelice:2010aj} to represent the equations  of the $f(R)$ gravity,  written for the physical metric $g_{\mu\nu}$ (Jordan frame), in the form of the usual Einstein equations for a conformally transformed metric $\hat g_{\mu\nu}$ (Einstein frame)
\begin{equation}\label{conformal_xi}
 g_{\mu\nu} = e^{-2\xi} \hat g_{\mu\nu}\, , \quad  e^{2\xi}= f'(R)  \,, 
 \end{equation}
 accompanied by equations for an additional canonically normalized  scalar field (SF) $\phi = \sqrt{6/ \kappa}\, \xi$.  In this paper, we will describe this scalar field (the scalaron)  by the dimensionless variable $\xi$.

 The self-interaction potential $W(\xi)$ of the scalar field $ \xi$ can be introduced parametrically as follows:
\begin{equation} \label{SF-potential}
 e^{2\xi}=f'(u)\,,\quad 
 W(\xi)=\frac12  e^{-4\xi}\left[f(u)-f'(u)u\right]\,,
 \end{equation}
where the monotonicity of $f(u)$ is assumed.
The extrema of this potential are located at $\xi=\xi_m$ corresponding to $u=u_m$, which satisfy the equation 
\begin{equation}
    u_m f'(u_m) = 2 f (u_m) \, .
    \label{u_m}
\end{equation}
  
Denote
\begin{equation} \label{hatsTmunu}
\quad \hat T_{\mu\nu}= e^{-2\xi} T_{\mu\nu}\,.
\end{equation} 
To avoid confusion due to the presence of two metrics, we emphasize that, from this point on, ``hats'' correspond to the Einstein frame \cite{Sotiriou:2008rp, DeFelice:2010aj}, in which   covariant   differentiation, raising and lowering indexes are performed  with the metric tensor $\hat g_{\mu\nu}$.
On the account of  \eqref{conformal_xi} and \eqref{SF-potential}, equation (\ref{equations-f(R)}) leads to the following equations in the Einstein frame:  \begin{equation}\label{equations_EinFrame}
\hat R_{\mu\nu}-\frac12 \hat g_{\mu\nu}\hat R= 
  \hat   T^{(\xi)}_{\mu\nu}+  \kappa\hat T_{\mu\nu}  \,,
\end{equation}
where 
  \begin{equation} \label{SF-tensor}
	\hat T^{(\xi)}_{\mu \nu } =   6 \partial_{\mu} \xi\, \partial_{\nu}\xi - \hat g_{\mu \nu } \left[3 \xi_{,\alpha } \xi_{,\beta } \hat g^{\alpha \beta } -W(\xi )\right]  \,.
	 \end{equation}
Equations (\ref{equations_EinFrame}) are supplemented by the  equation for the scalaron  $\xi$: 
\begin{equation} \label{equation_SF}
\hat{\nabla}_\alpha\hat{\nabla}^\alpha\xi=-\frac{1}{6}\frac{\partial W}{\partial\xi}+\frac16 \kappa   \hat   T \,,\quad \hat T=\hat g^{\mu\nu}\hat T_{\mu\nu}\,.
\end{equation}
The system of equations \eqref{equations_EinFrame}, \eqref{equation_SF} is  equivalent to the fourth-order equations (\ref{equations-f(R)}).

The quadratic part of $W(\xi)$ around its minimum determines the scalaron mass $\mu$\,:
\begin{equation}
 \frac16 W''(\xi_m)= \mu^2 = \frac13\left( \frac{u_m}{f'(u_m)} - \frac{1}{f''(u_m)}\right)  \, ,
 \label{mu^2}
\end{equation}
where  $u_m$ is determined by (\ref{u_m}). Non-observation \cite{Kapner:2006si, Adelberger:2006dh} of the scalaron-induced Yukawa forces \cite{Stelle:1977ry} between non-relativistic masses leads to a lower bound on the scalaron mass (see also \cite{Cembranos:2008gj, Cembranos:2015svp, Perivolaropoulos:2019vkb}):
\begin{equation}\label{mlow}
\mu \geq 2.7~\text{meV} \quad \text{at 95\% C.L.}
\end{equation}
This lower bound leads to a rather small  length scale $l_\mu=1/\mu$.

The covariant conservation law  (\ref{energy-momentum_conservation}) also must be rewritten in the Einstein frame. For the  energy-momentum tensor, we have
\begin{equation}\label{conservation_Einstein_frame} 
\hat \nabla_\mu \hat T^{ \mu}{}_{\nu}=  -   \hat T   \partial_\nu \xi \,,  \quad 
\hat T=\hat  g_{\alpha\beta}\hat T ^{\alpha\beta}  \,.   
\end{equation}

\section{Quadratic $f(R)$ gravity}
\label{quadratic f(R) static}
In what follows, we deal with the simplest model (\ref{R2}). 
 In the original inflationary model due to Starobinsky \cite{Starobinsky:1980te}, described by this Lagrangian, the scalaron mass has relatively large value $\mu \approx 3 \cdot 10^{13}$~GeV \cite{Planck:2018jri}. 

The scalaron potential \eqref{SF-potential} for model \eqref{R2} is 
\begin{equation}\label{SF-potential_exact}
 W(\xi)=\frac34 \mu^2 \left(1-e^{-2\xi}\right)^2  \,.
 \end{equation}
It is monotonically increasing with $\xi$ in the domain $\xi \geq 0$ asymptotically rising to a plateau. In the case of an  asymptotically flat space-time in the initial (Jordan) frame, we assume that the scalar curvature vanishes at spatial infinity; correspondingly, $\xi=\frac12 \ln[ f'(R)]\to 0$. 

For most of astrophysical systems with continuous distributions of matter,  we expect $\xi$ to be very small. 
For  $|\xi|\ll 1$, we have
\begin{equation}\label{SF-potential_approx}
W(\xi)   \approx 3 \mu^2  \xi^2\,.
\end{equation}  
In a static case, we deal with metric $g_{\mu\nu}\equiv g_{\mu\nu}({\boldsymbol r})$;  $g_{0i}\equiv0$, yielding   
\begin{equation}\label{equation_SF_approx}
\hat \Delta \xi\equiv- \frac{1}{\sqrt{-\hat g}}\frac{\partial}{\partial x^i}\left[\sqrt{-\hat g}\hat g^{ij}\frac{\partial \xi}{\partial x^i}\right]=\mu^2 \xi-\frac{\kappa}{6}\hat T\,, 
\end{equation}
where $i,j=1,2,3$.
 
As noted above, the value of the scalaron mass $\mu$ is bounded from below by (\ref{mlow}) so as not to contradict the  existing observations. A somewhat stronger lower bound, $\mu \gtrsim 4.4 $~meV, arises in the consideration of the scalaron as a dark-matter candidate \cite{Shtanov:2021uif, Shtanov:2022xew}. Such a mass corresponds to the length scale $l_{\mu} = \mu^{-1 } \lesssim 4.5\cdot 10^{-3}$~cm, which is a very small value in the astrophysical realm. In the case of an astrophysical object with mass $\mathfrak{M}$ and gravitational radius $r_g = 2 G \mathfrak{M}$,  we typically deal with a very large  dimensionless quantity $\mu r_g\gg 1$. This allows us to make a general estimate for the value of $\xi$ inside a sufficiently smooth distribution of $\hat T$. Indeed,  equation (\ref{equation_SF_approx}) can be written in a form that may be considered as a source of asymptotic approximations:  
\begin{equation} \label{equation_SF_asympt}
\xi=\frac{\kappa}{6\mu^2}  \hat T- \frac{1}{\mu^2}\hat \Delta \xi\,,
\end{equation}
with the last term on the right-hand side regarded as a small perturbation. This gives us a simple approximate  formula 
\begin{equation}\label{solution_SF_asympt_internal}
\xi\approx\frac{ \kappa}{6\mu^2} \left(   \hat T - \frac{1}{\mu^2}\hat \Delta \hat T \right)\,.
\end{equation} 
This formula is applicable if $ \mu^2 \hat T \gg  \hat \Delta \hat T$, which, however, may fail to be satisfied at the boundary of the body, where \eqref{solution_SF_asympt_internal} must be replaced by a more exact relation.

Further in this paper we focus primarily   on mathematical properties of the model and allow for arbitrary values of $\mu$ where possible.

\section{Spherically symmetric configurations}
 \label{SSS-Ein_frame}
 
\subsection{Field equations in the Einstein frame}\label{SSS-eqs}

The metric of a spherically symmetric space-time in the Schwarzschild (curvature) coordinates can be written as 
\begin{equation} \label{metric} 
d\hat s^{2} =e^{\alpha } dt^{2} -e^{\beta } dr^{2} -r^{2} \left(d\theta ^{2} +\sin ^{2} \theta d\varphi ^{2} \right) ,
\end{equation} 
where $r>0$, $\alpha\equiv \alpha(t,r)$, $\beta\equiv \beta(t,r)$. The nontrivial Einstein equations (\ref{equations_EinFrame}) in this case are 
\begin{align}
&\frac{\partial}{\partial r} \left[ r \left(e^{-\beta}-1\right) \right]=-\kappa r^2 \hat T_0^0 \nonumber \\ &\quad -
r^2 \left[ 3 e^{-\alpha}\left(\frac{\partial\xi}{\partial t}\right)^2+3 e^{-\beta}\left(\frac{\partial\xi}{\partial r}\right)^2+  W(\xi)\right] \,,
    \label{Ein_1-0} \\[3pt] 	
&{re^{-\beta}\frac{\partial\alpha}{\partial r}+e^{-\beta}-1} =-\kappa r^2 \hat T_1^1 \nonumber \\ &\quad + r^2 \left[3 e^{-\alpha}\left(\frac{\partial\xi}{\partial t}\right)^2+3 e^{-\beta}\left(\frac{\partial\xi}{\partial r}\right)^2-  W(\xi)\right]\,,
\label{Ein_2-0}
\end{align}
\begin{equation} 
\label{Ein_3-0}
\frac{\partial\beta}{\partial t}=- \kappa r e^\beta \hat T_0^1+6 r\frac{\partial\xi}{\partial t}\frac{\partial\xi}{\partial r} \,,
\end{equation}
where $\xi\equiv\xi(t,r)$, and the structure of $\hat T_{\mu\nu }$ corresponds to the spherically symmetric  case. 
 
Equation (\ref{equation_SF}) yields
\begin{align} 	
e^{- \frac{\alpha+\beta}{2}} \left[ \frac{\partial}{\partial t} \left( e^{\frac{\beta-\alpha}{2}} \frac{\partial  \xi}{\partial t} \right) - \frac{1}{r^2}  \frac{\partial}{\partial r} \left(r^2 e^{\frac{\alpha-\beta}{2}} \frac{\partial\xi}{\partial r} \right) \right] \nonumber\,\\= -\frac16 \left[W'(\xi) - {\kappa} \hat T \right] \,.
  \label{equation-phi}
 \end{align}
Note that, for  $\xi\equiv 0$ and $\hat T =0$, we get the Schwarzschild solution.

\subsection{Static SS solutions}
\label{asymptotics-of-static-solutions}

In the case of a static isolated regular configuration, the quantities $\alpha$, $\beta$ and $\xi$ depend only on $r$. We assume that,  for $r<R_0$, we have    a continuous matter distribution with non-zero regular  $\hat T_{\mu\nu}$, whereas $\hat T_{\mu\nu}=0$
in an outer region $r>R_0$. 

In what follows, we consider purely gravitational case with $\hat T=0$. For a static case, Eqs.~(\ref{Ein_1-0}),  (\ref{Ein_2-0})  yield 
\begin{align} \label{18-19} 
\frac{d}{dr}(\alpha+\beta) &= 6r\left(\frac{d\xi}{dr}\right)^2 \,, \\[3pt]        
\label{18+19} 
\frac{d}{dr}( \alpha-\beta) &= -\frac{2}{r}+\frac{2e^\beta}{r} \left[1-r^2W(\xi)\right]\,.
\end{align} 
The system of equations with respect to $\alpha$, $\beta$, and $\xi$ is closed by adding the equation for the SF: 
\begin{equation} 	
\label{equation-xi-vacuum}
\frac{d}{dr}\left[r^2 e^{\frac{\alpha-\beta}{2}}\frac{d\xi}{dr}\right]= \frac{r^2}{6} e^{\frac{\alpha+\beta}{2}} W'(\xi)   \,.
\end{equation}

In the case of an  asymptotically flat  static space-time, we assume 
  \begin{equation} \label{asympt-r-to-infty} 
\lim\limits_{r\to\infty} [ r\alpha(r)]=- r_{g}  ,\quad \lim\limits_{r\to\infty}[r\beta(r)]=r_{g} \,,   \end{equation} 
where $r_{g} =2G\mathfrak{M}$, and   $\mathfrak{M}>0$ is the configuration mass.

In the asymptotically flat configuration, we assume $\xi(r)\to 0$ as $r\to\infty$,   and Eq.~(\ref{equation-xi-vacuum}) can be approximated by the equation for a free massive scalar field on the Schwarzschild background \cite{Asanov1968, Asanov1974,Rowan1976}:
\begin{equation} 	
\label{equation-xi-small}
\frac{d}{dr}\left[r  \left(r-r_g\right)\frac{d\xi}{dr}\right]=  {r^2} \mu^2\xi   \,.
\end{equation}
We discard unbounded solutions of (\ref{equation-xi-small}) as $r \to \infty$ leaving only those with exponentially decaying $\xi(r)$.  
Taking into account the results  of \cite{Asanov1968, Asanov1974, Rowan1976, Stashko_Zhdanov_2019a} on the asymptotic behavior at infinity, we assume  
\begin{equation}\label{xi_inf}
\lim\limits_{r\to\infty}\left[\left(\frac{r}{r_g}\right)^{1+ M\mu}e^{\mu r}\xi(r)\right]=Q\,. 
\end{equation} 
Here and below, $M=r_g/2$, and the constant $Q$ measures the strength of the scalaron field at spatial infinity and will be dubbed as the ``scalar charge." 

For given $\mu>0$, $\mathfrak{M}>0$ and $Q$, we  claim  that  \textit{there is  $r_0>R_0$ such that solution $\alpha(r)\in C^1$, $\beta(r)\in C^1$, $\xi(r)\in C^2$   of  Eqs.~(\ref{18-19})--(\ref{equation-xi-vacuum})  exists for $r\ge r_0$ and is uniquely defined by conditions (\ref{asympt-r-to-infty}), (\ref{xi_inf})} regardless of the interior solution and structure of the energy-momentum tensor. This statement is physically quite understandable; however, its rigorous proof requires some analytical work, which we present in Appendix \ref{r_to_infty}.

Our next step is to show that, if  $\hat T_{\mu\nu}\equiv 0$  for all $r>0$, then one can put $r_0=0$;  in this case, the parameters $M$ and $Q$ completely define the static  SS  configuration. This is the subject of the next Section \ref{sec:asympt}.

The numerical investigations, which we perform below, require more detailed information on the asymptotic properties at large $r$. This  is also considered in Appendix~\ref{r_to_infty}, where we present a convergent approximation method to justify  asymptotic of decaying static solutions $\xi(r)$  for $\mu r\gg1$, $r\gg r_g$. The leading terms of this asymptotics for $|\xi|\ll 1$ are
\begin{equation} \label{asy-phi-infty} 
\xi(r)=Q \left(\frac{r_g}{r}\right)^{1+ M\mu} \left(1+\frac{b_1}{r}+\frac{b_2}{r^2}\right) e^{-\mu r}\, ,   \end{equation} 
where 
\begin{align}
b_1 &= \frac{M}{2}(1+3 M\mu)\,, \\[3pt]
b_2 &= \frac{M}{8 \mu} (2 + 5 M \mu + 16 M^2 \mu^2 + 9 M^3 \mu^3)\,. 
\end{align}
Note that (\ref{asy-phi-infty}) differs by a power-law factor from the usual  Yukawa asymptotics in flat space. In view of the consideration in Appendix \ref{r_to_infty}, an  asymptotic relation for $d\xi/dr$ can be obtained by formal differentiation of (\ref{asy-phi-infty}).
 
\section{Global behavior of static solutions and asymptotic properties near NS}\label{sec:asympt}

Now we consider a purely gravitational SS system, that is $\hat T_{\mu\nu}\equiv 0$. {Transition to the Einstein frame  enables us to use the earlier results on the global  properties of the SS systems with a scalar field. In this section, we  use the method of \cite{ZhdSt}  with minor modification.}

According to the previous results, there exists an asymptotically flat  solution of Eqs.~(\ref{18-19})--(\ref{equation-xi-vacuum})     on $[r_0,\infty)$ for some $0<r_0<\infty$. {The set of all such $r_0$ has} infimum $r_0^*=\inf{r_0}<\infty$, and the solution exists for $r\in (r_0^*, \infty)$. {We prove that $r_0^* = 0$.}

Suppose (on the contrary) that $r_0^*>0$ and consider the solution on $ (r_0^*, \infty)$. 

The SF potential (\ref{SF-potential_exact}) has the property
\begin{equation}
\xi W'(\xi)>0,\quad \xi\ne 0\,.
\label{monotonocity}
\end{equation}
Using this equation, it is easy to see that
\begin{equation} 	\label{sign-xi-vacuum}
 	\frac{d\Gamma}{dr}=r^2 e^{\frac{\alpha-\beta}{2}}  \xi'^2 + \frac{r^2}{6} e^{\frac{\alpha+\beta}{2}}\xi W'(\xi)   \,,
 \end{equation}
where 
$
\Gamma(r)\equiv r^2 e^{\frac{\alpha-\beta}{2}}\xi\, \xi'$, $\xi'\equiv  d\xi/dr$.

The right-hand side of (\ref{sign-xi-vacuum}) is positive for any  nontrivial  $\xi(r)$, and the function $\Gamma(r)$ is strictly increasing. For the solutions satisfying (\ref{xi_inf}), we have $\Gamma(r)\to 0$ as $r\to \infty$.  In a nontrivial case, this is possible if and only if $\Gamma(r)<0$, so that \textit{ $\forall r>r_0^*$ the functions $\xi(r)$ and  $\xi'(r)$ are nonzero and have opposite signs}.

Now we will prove that, in the case of  $\hat T_{\mu\nu}\equiv 0$, solutions of  (\ref{18-19})--(\ref{equation-xi-vacuum})  {with asymptotic conditions \eqref{asympt-r-to-infty}, \eqref{xi_inf}} 
can be regularly extended to the left of $r_0^*$.

Similar to \cite{ZhdSt}, we introduce  variables  $x=(\alpha+\beta)/2$ and
\begin{equation}
X = e^x \,, \quad Y=\left(\frac{r}{r_g}\right) e^{\frac{\alpha-\beta}{2}} \,, \quad Z=-rY\frac{d\xi}{dr} \, .
\end{equation} 
Then, using (\ref{18+19}) and (\ref{18-19}), we obtain an equivalent first-order system in terms of the dimensionless variables $X$, $Y$, $Z$, and $\xi$\,: 
\begin{subequations}\label{Ein_system}
\begin{align}
\frac{dX}{d r} &=\frac{3XZ^2}{rY^2}\,,  \label{Ein_system-A} \\
\frac{d Y}{d r} &=\frac{X}{r_g} \left[1-   r^2 W(\xi) \right]\,,  \label{Ein_system-B} \\
\frac{d Z}{d r} &=- \frac{r^2}{6r_g} X \,W'(\xi)\,, \label{Ein_system-C}\\    
\frac{d \xi}{d r} &=- \frac{Z}{rY}  \,, \label{Ein_system-D} 
  \end{align}
\end{subequations} 
It is sufficient to show that the right-hand sides of system (\ref{Ein_system}) are bounded as $r\to r_0^*$. We shall  use the monotonicity properties of $X(r)$, $Z(r)$, and $\xi(r)$ following directly from (\ref{Ein_system-A}), (\ref{Ein_system-C}), (\ref{Ein_system-D}).

Consider first the case $\xi(r)>0$. In the  domain $\xi>0$, where $W(\xi)$ and $W'(\xi)$ are bounded,  Eq.~(\ref{Ein_system-A}) implies that $X(r)>0$ is monotonically increasing and bounded for $r\to r_0^*$. 
Therefore, the right-hand sides of (\ref{Ein_system-B}),(\ref{Ein_system-C}) are bounded and integrable as $r\to r_0^*+0$ and there exist finite limits $Z(r)\to Z_0$ and $Y(r)\to Y_0$.  

Evidently, $Z _0 >0$ because $Z(r)$ is a strictly decreasing function as a consequence  of    (\ref{Ein_system-C}). 

In order to estimate the value of  $Y_0$ from below, we consider an  interval $(r_0^*,r_1]$ for some $r_1<\infty$. Taking into account that  $X(r)$ is a monotonically increasing function,  we can choose it as an independent variable. After division of  (\ref{Ein_system-B}) by  (\ref{Ein_system-A}), we have, for $r_0^*<r<r_1$,
\begin{equation*}
-\frac{d}{dX}\left(\frac{r_g}{Y}\right)= \frac{r}{3Z^2(r)}\left[1- r^2 W(\xi(r))\right]< \frac{r_1}{3Z^2(r_1)}\,,
\end{equation*}
where we have taken into account $Z(r)>Z(r_1)$. After integration from $X(r)$ to $X(r_1)$, we obtain
\begin{equation}\label{esteem_1/Y}
\frac{1}{Y(r)} < \frac{1}{Y(r_1)}+ \frac{r_1 X(r_1)}{3r_gZ^2(r_1)}\,.
\end{equation}
This excludes the case $Y_0=0$ yielding $Y_0>0$. 

Now we see that  the right-hand sides of (\ref{Ein_system-A}),(\ref{Ein_system-D}) are also  bounded and integrable and there exist finite limits $X(r)$ and $\xi(r)$ for  $r\to r_0^*>0$.

Thus, the whole  system (\ref{Ein_system}) 
is regular for $r\to r_0^*$  and, according to the local existence and uniqueness theorems, it can be extended to the left neighbourhood of this point. The  contrary assumption is false and we must put $r_0^*=0$.  

Now we can repeat considerations concerning $Y(r), Z(r)$ on $(0,\infty)$ yielding 
\begin{equation}\label{limit_z}  Z_0=\lim\limits_{r\to 0^+} Z(r) >0   
\end{equation} 
and
\begin{equation} \label{limit_Y} 
Y_0=\lim\limits_{r\to 0^+} Y(r) > 0\,.        
\end{equation}
This completes the consideration of the case $\xi(r)>0$.

The case of $\xi(r)<0$ differs from that considered above due to the exponential behavior of $W(\xi)$, $W'(\xi)$ for $\xi<0$.
However, here we also can show that the right-hand side of (\ref{Ein_system-C}) is bounded  by  using the same reasoning as in \cite{ZhdSt} (Lemma~4); see Appendix \ref{xi<0}.
Further  consideration is similar to the case of positive $\xi(r)$. 

Finally, we summarize  that \textit{solution $\alpha(r)\in C^1$, $\beta(r)\in C^1$, $\xi(r)\in C^2$   of  Eqs.~(\ref{18-19})--(\ref{equation-xi-vacuum})    satisfying (\ref{asympt-r-to-infty}), (\ref{xi_inf}) exists for all $r> 0$  and is unique}. 
\textit{Moreover, there exist the limits from the right  (\ref{limit_z}), (\ref{limit_Y})}.

Using the estimates of $Y_0$ and $Z_0$, we infer a  logarithmic behavior of $\alpha$, $\beta$ and $\xi$ corresponding to a power-law behavior of $X(r)$ and $Y(r)$  as $r\to 0$:  
\begin{equation} \label{asympt_chi_xi}
x\sim  \eta\, \ln \left(\frac{r}{r_g}\right) \,,\quad \xi (r)\sim -\zeta\, \ln\left( \frac{r}{r_g}\right) \,,  
\end{equation}
\begin{equation} \label{asympt_alpha_beta} 
 \alpha (r)\sim (\eta -1)\, \ln\left( \frac{r}{r_g}\right) \, ,  \quad   \beta (r)\sim (\eta +1)\, \ln\left( \frac{r}{r_g}\right) \,,        
\end{equation}
where $\zeta=Z_0/Y_0>0$ and $\eta =3 \zeta ^{2}$. These constants can be related with asymptotic parameters $\mathfrak{M}$ and $Q$ at infinity  by means of   numerical methods. 

Relations (\ref{asympt_chi_xi}), (\ref{asympt_alpha_beta}) justify the choice of zero approximation for the iteration procedure described in Appendix 
\ref{Near_NS}, which enables us to obtain more detailed  
asymptotic relations for $r\to 0$. 
Here we present  the resulting  leading orders of the metric coefficients,  
\begin{equation} \label{solution_as_ea} 
e^\alpha(r)\approx {X_0Y_0} \left(\frac{r}{r_g}\right)^{\eta-1}\left[ 1-\sigma\frac{\eta-1}{(\eta+1)^2}\left(\frac{r}{r_g}\right)^{\eta+1}\right]
\,,     
\end{equation}
\begin{equation} \label{solution_as_eb} 
e^\beta(r)\approx\sigma\left(\frac{r}{r_g}\right)^{\eta+1} \left[ 1-\sigma\frac{3\eta+1}{(\eta+1)^2} \left(\frac{r}{r_g}\right)^{\eta+1}\right]
\,,     
\end{equation}
and of the scalaron field,
\begin{equation} \label{solution_as_xi} 
\xi(r)\approx-\zeta\, \ln \left(\frac{r}{r_g}\right) +\xi_0+\frac{\zeta\sigma}{(\eta+1)^2}\left(\frac{r}{r_g}\right)^{\eta+1}\,,
\end{equation}
where $\zeta=Z_0/Y_0,\,\sigma=X_0/Y_0$, $\eta =3 \zeta ^{2}$, and $X_0>0$ is the   constant arising in (\ref{chi-xi_int_a}). In case of asymptotically flat configurations these constants are related to the parameters $Q$ and $M\mu$. These dependencies can be obtained numerically; the example for $\eta=\eta(M\mu)$ is presented on Fig. \ref{p3.png}.

{Obviously, our conclusion about the  qualitative behavior of the asymptotically flat solutions on the radial interval $(0,\infty)$  is  valid in the Jordan frame as well. Some caution is needed about the singularity at the origin.  Similarly to \cite{ZhdSt},   using the asymptotic relations  (\ref{solution_as_ea}), (\ref{solution_as_eb}), we infer that there is a naked singularity at $r=0$ in the Einstein frame. However, due to the singular conformal factor (\ref{conformal_xi}), we must check the appearance of NS directly in the Jordan frame (cf.\@ \cite{Bahamonde-2016}).} 
For the original metric in the Jordan frame, we have 
\begin{equation}\label{metric_Jordan}
ds^2=A(r)dt^2-B(r)dr^2 -\rho^2(r)\left(d\theta ^{2} +\sin ^{2} \theta d\varphi ^{2} \right) \, ,
\end{equation}
where according to (\ref{conformal_xi})
 \begin{equation} \label{asympt_JordanA} 
A(r)=e^{\alpha-2\xi}\,,\quad  B(r)=e^{\beta-2\xi}\,,\quad \rho(r)=re^{-\xi} \,.
 \end{equation}
The Kretschmann scalar behaves as
 \begin{equation} \label{Kretschmann} 
K=R_{\alpha \beta\gamma\delta}R^{\alpha \beta\gamma\delta}\sim \left(\frac{r_g}{r}\right)^{6(1+\zeta^2)+4\zeta}\,.
 \end{equation}
Using (\ref{asympt_JordanA}), it is easy to verify that radial null geodesics emanating from the origin can reach an external observer in a finite time. Thus, for  arbitrarily small SF $\xi\ne 0$, $Q\ne 0$, there is a naked singularity at the center.\footnote{Here, the center of the spherically symmetric system is  defined as the origin of the curvature coordinates for metric (\ref{metric}).}    

Transition from (\ref{metric_Jordan}) to the curvature coordinates $t,\rho,\theta,\varphi$ yields
\begin{equation}\label{metric_Jord_curv}
ds^2=\tilde{A}dt^2-\tilde{B}d\rho^2 -\rho^2\left(d\theta ^{2} +\sin ^{2} \theta d\varphi ^{2} \right)\, ,
\end{equation}
where the leading terms as $\rho\to 0$ are 
 \begin{align*} \label{asympt_Jord_curv} 
\tilde{A}(\rho)=&  {X_0Y_0} e^{3\xi_0(\zeta-1)}\left(\frac{\rho}{r_g}\right)^{t}\left[1+ {\cal O} \left(\frac{\rho}{r_g}\right)^s\right]\,,\quad \\ \tilde{B}(\rho)=& \frac{\sigma}{(\zeta+1)^2}e^{s\xi_0}\left(\frac{\rho}{r_g}\right)^s \left[1+ {\cal O} \left(\frac{\rho}{r_g}\right)^s\right]\,,
 \end{align*}
 \[
t= {3\zeta-1} \,,\quad s=\frac{3\zeta^2+1}{\zeta+1}\,,\quad \zeta\ne -1\,.
  \]
The integer values of $s,t$ found in \cite{Hernandez2020} can  be obtained in part with $\zeta=-2/3, \pm 1/3, 1, 3$.  The cases $(t, s)=(0, 0),(1,-1)$ indicated in \cite {Hernandez2020} turn out to be impossible in our analysis of asymptotically flat configurations with $\zeta\ne 0$. Note that taking into account the next order terms  from (\ref{solution_as_ea})--(\ref{solution_as_xi}) shows that metric coefficients in the general case cannot be represented in the form of the Frobenius expansion, as assumed in \cite {Hernandez2020} (cf. also \cite{Lu-Stelle-2025}); this is  true only for special values of $\zeta$.

\section{Spherical perturbations}\label{SS_perturbations}
Our aim  is to find a region of parameters $\mu, Q$ for which static SS configurations described by Eqs.~(\ref{Ein_1-0})--(\ref{equation-phi}) with $\hat T_{\mu,\nu}\equiv 0$ are unstable against small perturbations.  
To check for stability/instability issues, we consider linear    perturbations, which can be expressed  as superposition of functions with the time dependence\footnote{Strictly speaking,   we consider a class of time-dependent perturbations that satisfy certain boundary conditions as functions of $r$ (see below) and can be expressed in the form of a Laplace/Fourier transform as functions of $t$. } $\sim \exp(-i\Omega t)$,  $\Omega\ne 0$. The boundary conditions for the spatial dependence of the perturbations will be given below.    
  
 There is an extensive literature on the linear  perturbations against spherically symmetric background  (see, e.g., \cite{Chandrasekhar1998, Berti2009}). The perturbations can be separated into axial and polar modes  \cite{Berti2009,Chandrasekhar1998}, which can be treated independently. However, in order to show that the system is unstable, it is sufficient to show that there exists at least one unstable mode.  Correspondingly, we will restrict ourselves to the case of radial perturbations. 
  In this case, the treatment of our problem follows the same scheme as in  \cite{CLAYTON1998131}.  
  
Based on the exact  equations (\ref{Ein_1-0})--(\ref{equation-phi}) in vacuum ($\hat T_{\mu \nu}\equiv 0$), we perturb the static SS  background  $\alpha(r)$, $\beta(r)$, $\xi(r)$ by considering   $\alpha(r)+\alpha_1(t,r)$, $\beta(r)+\beta_1(t,r)$, $\xi(r)+\xi_1(t,r)$, where  $\alpha_1$, $\beta_1$, $\xi_1$ represent small perturbations.

After linearization, Eq.~(\ref{Ein_3-0})  yields 
\begin{equation*}
    \dot \beta_1= 6 r \dot \xi_1 \xi' \, . 
\end{equation*}
Neglecting the static additive, we have
\begin{equation}\label{beta_1}
      \beta_1= 6 r  \xi_1 \xi'  \,.
\end{equation}
The linearized sum of Eqs.~(\ref{Ein_1-0})--(\ref{Ein_2-0})  yields 
 \begin{equation}
\frac{\alpha_1'-\beta_1'}{2}=\xi_1 e^\beta\left\{6    \xi' \left[1-r^2W(\xi)\right] -{r}W'(\xi) \right\}\, . \label{sum} 
  \end{equation}
Eqs.~(\ref{beta_1}), (\ref{sum}) allow us to express all perturbations through $\xi_1$. Then we perform linearization of (\ref{equation-phi})    taking into account  this equation for the background values, and substitute $\xi_1 =r^{-1}\Phi$  to obtain the master equation in the form
\begin{equation}
\frac{\partial^2\Phi}{\partial t ^2}  - e^{\frac{\alpha-\beta}{2}}\frac{\partial }{\partial r}\left[ e^{\frac{\alpha-\beta}{2}} \frac{\partial \Phi}{\partial r}\right]+ U_{\rm eff} \Phi=0\,,
\label{master-radial0_T}
 \end{equation} 
 \begin{eqnarray} 	
U_{\rm eff}(r) =\frac{e^{\alpha-\beta}}{r} \frac{\alpha'-\beta'}{2}+  \frac16 e^{\alpha} W''(\xi) \qquad\qquad   \nonumber \\   -e^{\alpha}\xi' \left[ 6\xi'\left(1-r^2W(\xi)\right) -2rW'(\xi)   \right]                  \,. \quad 
   \end{eqnarray}
here $\alpha\equiv\alpha(r)$, $\beta\equiv \beta(r)$, $\xi\equiv\xi(r)$.

\begin{figure}
    \centering
    \includegraphics[width=.49\textwidth]{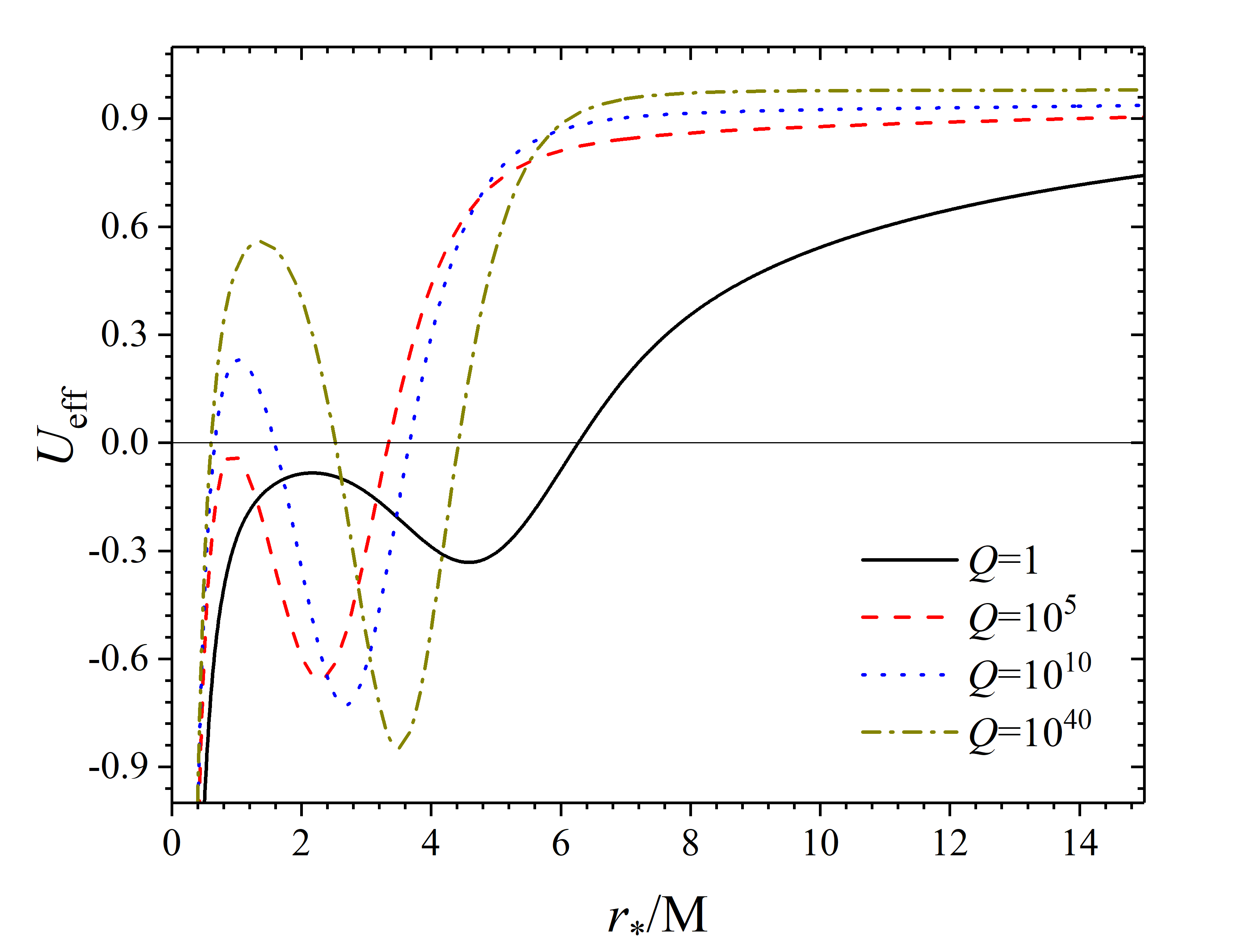}
    \includegraphics[width=.49\textwidth]{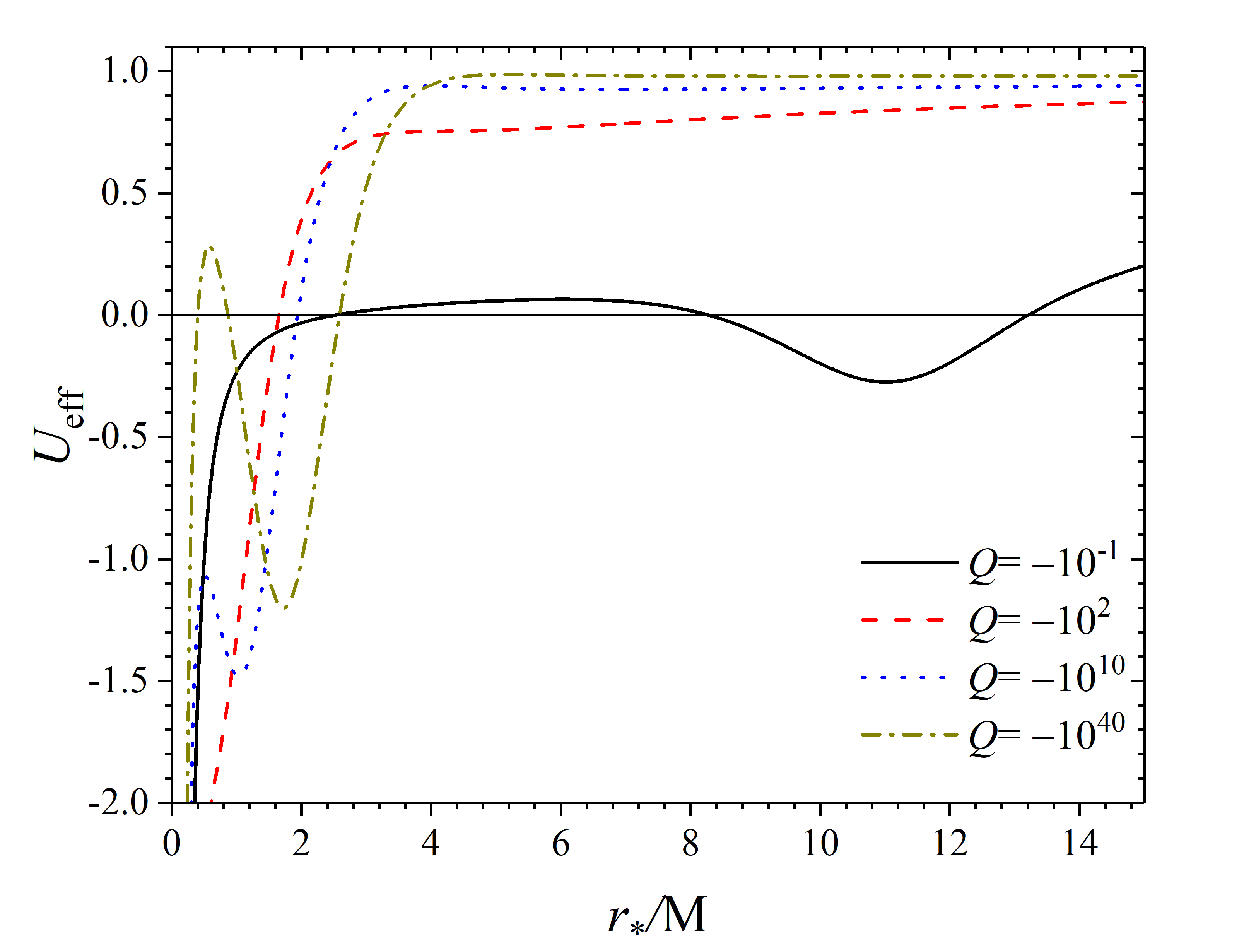}
    \caption{Typical examples of the effective potential $U_{\rm eff}[r(r_*)]$  for radial perturbations in  the case of  $M\mu=1$. The upper/lower panel shows the case of  positive/negative $Q$.  In both cases, $U_{\rm eff}\to\mu^2$ as $r_* \to\infty$.}
    \label{fig_eff_pot_rad}
\end{figure}

Using (\ref{18+19})   and (\ref{Ein_system-D}), we get
 \begin{eqnarray} 	
 U_{\rm eff}(r)=-\frac{e^{\alpha-\beta}}{r^2}+\frac{e^{\alpha}}{r^2}\left[1-r^2W(\xi)\right]\left( 1-6\frac{Z^2}{Y^2}\right) \nonumber \\ {} -2e^{\alpha}\left[\frac{Z}{Y} W'(\xi)-\frac{1}{12}W''(\xi)\right] \,. \quad 
  \label{equation-phi_p}
 \end{eqnarray} 
 Typical examples of $U_{\rm eff}$  are shown  in Fig. \ref{fig_eff_pot_rad}.

The initial value problem related to equation (\ref{master-radial0_T}) would be incomplete without a boundary condition at the center and at the infinity. While behavior at infinity is treated in a standard way,  we have no physical idea of what happens near a naked singularity. On the other hand, the problem must be mathematically correct in order to provide a unique solution to the linearized equations \cite{Wald1980,Who_afraids, CLAYTON1998131, Ishibashi-Wald1999,Gibbons2005,Dotti_2022}. Our analysis below is analogous to that of \cite{CLAYTON1998131}, which takes into account that we deal with small  perturbations.

For a single mode  $\Phi\sim \exp(-i\Omega t)$ we have from (\ref{master-radial0_T})
 \begin{equation}
   e^{\frac{\alpha-\beta}{2}}\frac{\partial }{\partial r}\left[ e^{\frac{\alpha-\beta}{2}} \frac{\partial \Phi}{\partial r}\right]+\left[\Omega^2-U_{\rm eff}\right]\Phi=0\,,
\label{master-radial0}
 \end{equation}

The asymptotic behavior of $U_{\rm eff}(r)$ as $r\to 0$
is dominated by the  term 
\begin{equation}
\frac{e^{\alpha-\beta}}{r^2} \approx \frac{K_0}{r^4}\left[1+ \frac{2\sigma}{\eta+1}\left(\frac{r}{r_g}\right)^{\eta+1}\right]\,, \quad 
\end{equation}
where $K_0= { Y_0^2r_g^2} $ 
 is not contained in  the general asymptotic solution of (\ref{master-radial0}) 
\begin{equation}
\Phi(r)= C_1\, r + C_2\, r\ln (r/r_g) \,,\quad  r\to 0\,.
\label{master-asymptotic}
\end{equation}
In order that this solution and its  derivatives be regular at the origin, we must put $C_2=0$; otherwise, the perturbation cannot be considered small. Moreover, if $C_2\ne 0$, then the perturbed Kretschmann scalar turns out to be much larger than (\ref{Kretschmann}) as $r\to 0$ (cf.\@ considerations in  \cite{Gibbons2005}). 

Therefore, we consider solutions of (\ref{master-radial0}) with regular first derivative at $r=0$ corresponding to 
\begin{equation}
\Phi(r)= C_1\, r \,, \quad C_2=0\,,\quad  r\to 0\,.
\label{master-solution_small_r}
\end{equation}
The boundary condition at spatial infinity is 
\begin{equation}
\Phi(r)\to 0 , \quad r\to \infty\,.
\label{Phi_at_infty}
\end{equation}

In terms of the 
``tortoise'' radial variable 
  $r_{*}$   defined as  
 \begin{equation}\frac{dr_*}{dr}=e^{(\beta-\alpha)/2} \,,\quad r_*(0)=0\,,
 \end{equation}
Eq.~(\ref{master-radial0}) can be written in the familiar form 
  \begin{equation}
     \frac{\partial^2 \Phi}{\partial r_*^2}-U_{\rm eff} \left[ r \left( r_* \right) \right]\,\Phi=-\Omega^2 \Phi\,.
\label{master-radial}
 \end{equation}

Thus we have the problem of finding an eigenpair $(\Omega^2, \Phi\in C^2)$ for the symmetric operator on the left-hand side of (\ref{master-radial}) with the null Dirichlet boundary condition at the center and condition (\ref{Phi_at_infty}) at infinity\footnote{For numerical investigations, it is more convenient to deal directly with initial equation (\ref{master-radial0}).}. To be precise,  we work in the  space of  $C^2$--functions, which are  square integrable on $(0,\infty)$.  Obviously, the eigenvalues $\Omega^2$ of this problem are real; for brevity, we call solutions ``unstable" if $\Omega^2<0$ and ``stable" otherwise.

For large values of the radial variable, taking into account (\ref{asy-phi-infty}) and (\ref{SF-potential_approx}),  we have 
\begin{equation}
    U_{\rm eff}(r)\approx \left(1-\frac{r_g}{r}\right) \mu^2\,,\quad r\to\infty.
\end{equation}
The corresponding asymptotic for large $r$ is 
\[
\Phi(r)\sim D_1(\Omega) e^{kr}+D_2(\Omega) e^{-kr}\,,\quad k=\sqrt{\mu^2-\Omega^2}\,.
\]
Therefore, a necessary  condition for (\ref{Phi_at_infty}) to be satisfied is $\mu^2>\Omega^2$.  
The equation for the  eigenvalues is  
\begin{equation}
D_1(\Omega)=0\,,
\label{eigenvalue}
\end{equation}
and it ensures condition (\ref{Phi_at_infty}).

\section{Numerical solutions}
\label{num}

\subsection{Static SS solutions}
\label{num_static}
  
We performed backward numerical integration of (\ref{18+19}), (\ref{equation-xi-vacuum}) with respect to  $\alpha$, $\beta$, and $\xi$  starting from some sufficiently large value  of the radial variable $r_\text{init}\gg r_g$ and $\mu r_\text{init} \gg 1$, where $\xi$ can be assumed sufficiently small and one can use the asymptotic relations (\ref{asy-phi-infty}) to specify the initial conditions.\footnote{Asymptotic formulas for $d\xi/dr$ are obtained by formal differentiation of (\ref{asy-phi-infty}).} We typically opted     $r_\text{init}\sim 100 r_g$ for moderate $\mu$. Note that, in this case, the backward integration is more convenient than using the shooting method, which involves several shooting parameters.

Numerical investigations show that $e^{\alpha(r)}$  is monotonically increasing, while there is a maximum of $e^{\beta(r)}$ at some $r_\text{max}$ so that this function is monotonically increasing in the interval $(0,r_\text{max})$ and decreasing for $r>r_\text{max}$. Near the origin, $e^{\alpha(r)}>e^{\beta(r)}$, but, at some $r = r_c$, the sign of the inequality changes and we have  $e^{\alpha(r)}<1< e^{\beta(r)}$ for $r>r_c$.

 For fixed $M,Q$, as $\mu$ increases, the maximum  of $e^{\beta(r)}$  grows and becomes very sharp, whereas, to the left of this maximum, the graphs of $e^{\alpha(r)}$, $e^{\beta(r)}$  are  pressed to the abscissa axis. The same situation is observed with fixed $\mu$ as $Q$ decreases and becomes small. 
To the right of the maximum, we have $e^{\alpha(r)}< e^{\beta(r)}$; as $r\to\infty$, these functions  tend  to the asymptotic value  equal to unity.  The larger is $\mu$, the faster the asymptotic values are reached. 
 
There are different modes of behavior of $Y(r)$. For fixed $M$ and $\mu$, and for a sufficiently small  $Q$, we have  $r^2 W(\xi(r))\le 1$ for all $r$, and $Y(r)$ is monotonically increasing  with the asymptotics $Y(r)\approx r$   for large $r$. 
For sufficiently large $|Q|$, we have an interval with negative right-hand side of (\ref{Ein_system-B}), where  $Y(r)$   is a decreasing function. This interval is bounded because $\xi(r)$ is an exponentially  decreasing function and, for sufficiently large $r$, again $r^2 W(\xi(r))<1$, and $Y(r)$ is monotonically increasing. This behavior is typical in case of large $M\mu$; in this case we have a plateau of almost constant $Y(r),Z(r)$ near the center.  
Far from the center, SF tends to zero exponentially according to Eq.~(\ref{asy-phi-infty}) and (\ref{phiWKB-N}) leading to $e^\alpha\approx 1-r_g/r\approx e^{-\beta}$.

For  $M\mu\gg 1$  we used the approximation (\ref{phiWKB-N}) of Appendix \ref{approximations-large_mu}, which enables us to get the initial conditions for the backwards numerical integration with, e.g.,  $r_\text{init}\sim (5\div10)r_g$  and  $\xi(r_\text{init})\sim 0.01$. 
Here we practically have  the Schwarzschild metric for $r>r_0$. Note that the case $M\mu\gg 1$ is quite difficult to process on a laptop due to the rapid change of  solutions near the region where the  condition of smallness of $\xi$ begins to be violated.
   
\begin{figure}[h!]
 	\centering
\includegraphics[width=0.49\textwidth]{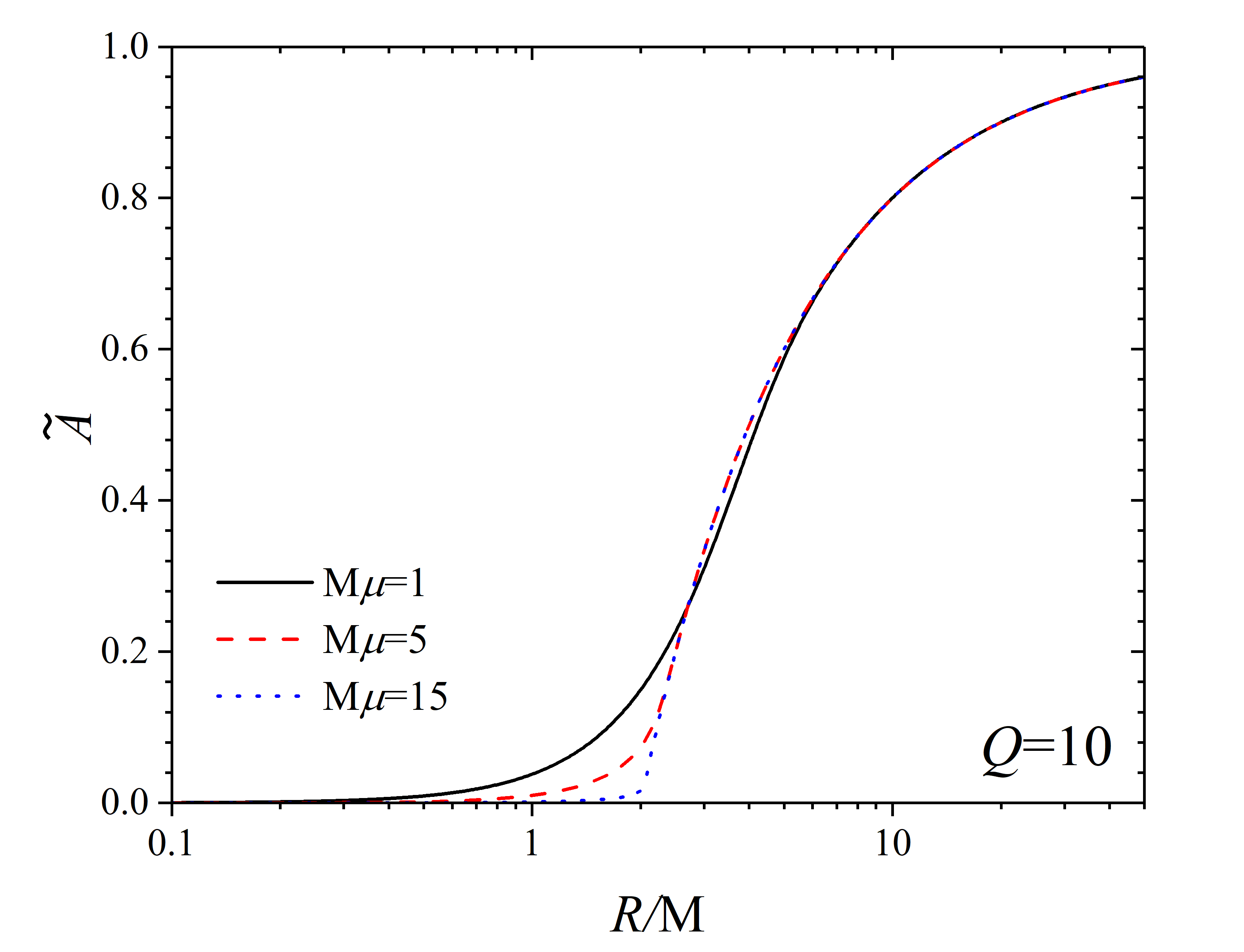}
 	\caption{Typical behavior of $\tilde{A}(\rho)$ in (\ref{metric_Jord_curv})  for different values of $M\mu$ and for $Q=10$.}
 	\label{fig:metricJordanA}
 \end{figure}

\begin{figure}[h!]
 	\centering
 \includegraphics[width=0.49\textwidth]{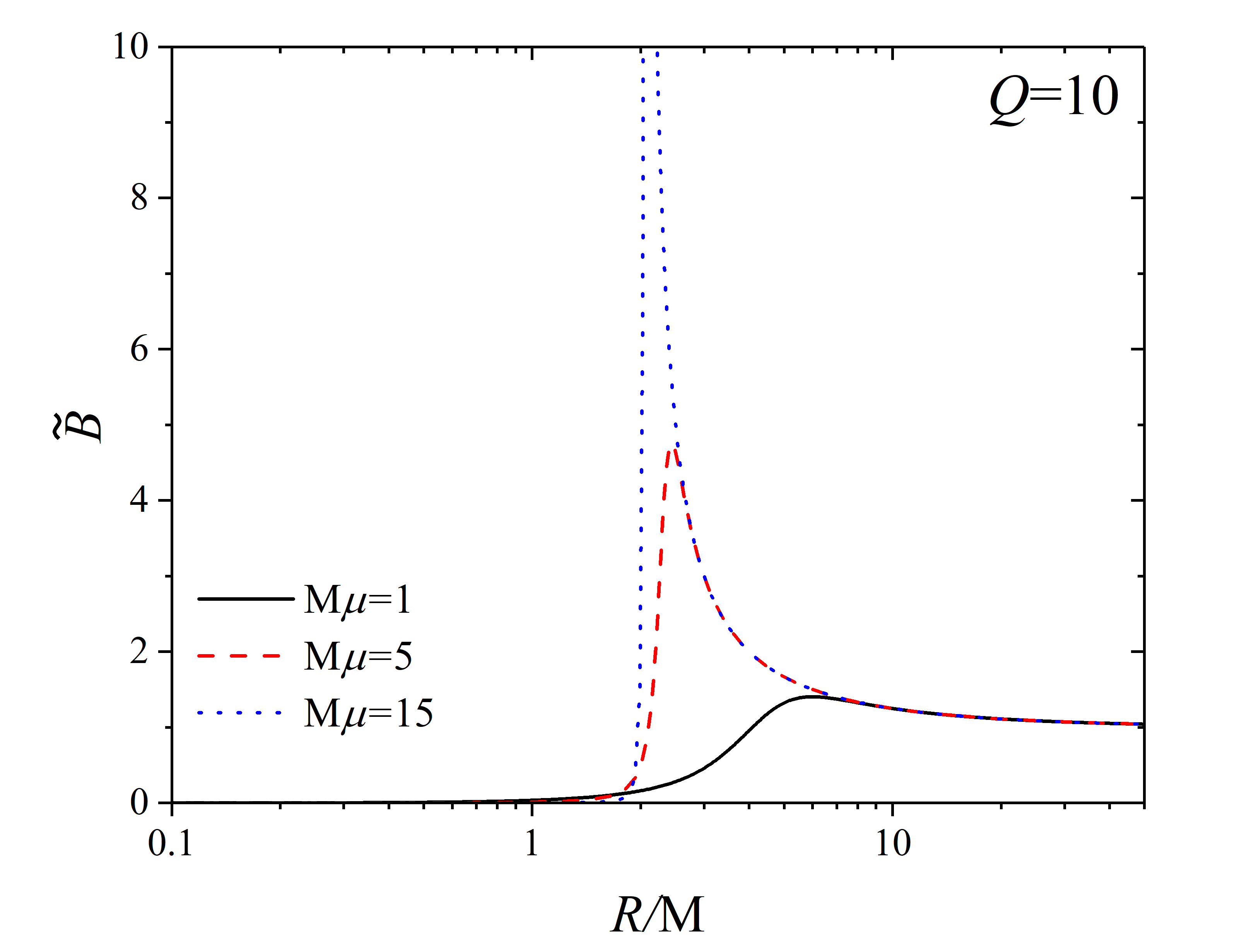}
 	\caption{Typical behavior of $\tilde{B}(\rho)$ in (\ref{metric_Jord_curv}) for different values of $M\mu$ and for $Q=10$.}
 	\label{fig:metricJordanB}
 \end{figure}

\begin{figure}[h!]
 	\centering
\includegraphics[width=.45\textwidth]{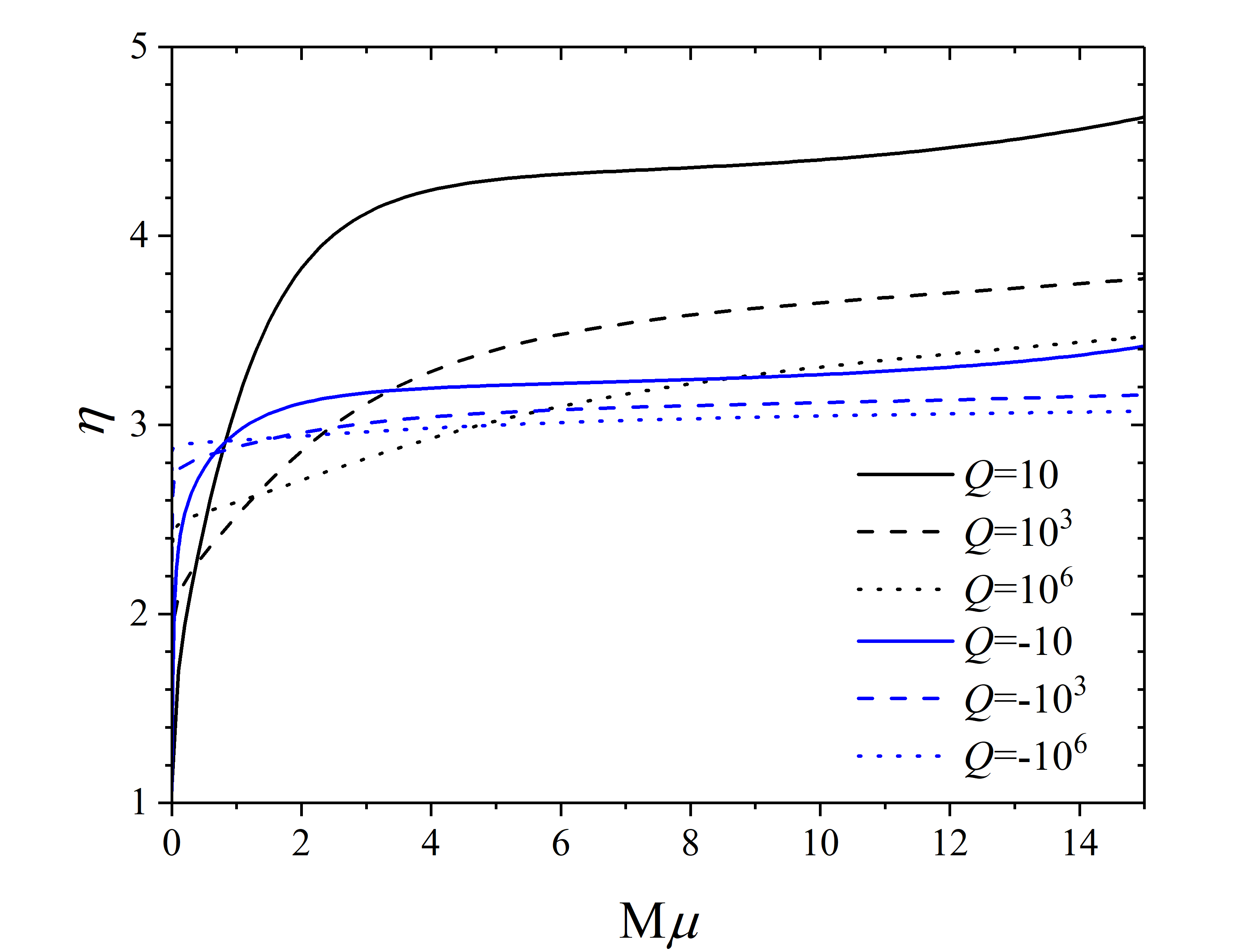}
 	\caption{Behavior of $\eta=3\zeta^2$ as a function of $M\mu$ for  different values of $Q$. }
 	\label{p3.png}
 \end{figure}

Figs.~\ref{fig:metricJordanA} and \ref{fig:metricJordanB} show the resulting metric coefficients $\tilde A(R)$ and $\tilde B(R)$  recalculated to the curvature coordinates of the Jordan frame (\ref{metric_Jord_curv}). Their behavior is qualitatively similar to the case of the Einstein frame. 
The values $(Y_0, Z_0)$ are derived numerically according to (\ref{limit_Y}), (\ref{limit_z}); this enables us to obtain the asymptotic parameters $\zeta$ and $\eta=3\zeta^2$. Figure \ref{p3.png} shows the dependence $\eta(M\mu)$. 

\subsection{(In)stability against radial perturbations}
\label{instability_radial}

Here we present the results of numerical analysis on linear stability/instability.
We imposed null Dirichlet conditions in the center, because otherwise perturbations cannon be considered small.
The algorithm of calculations involved two main stages.
\begin{itemize}

\item We chose the initial conditions for static SS (background) solutions  according to the asymptotic formula (\ref{asy-phi-infty}) at sufficiently large $r$ and perform backward integration to get these solutions according to subsection \ref{num_static}. 
 \item In order to check stability of the static SS solution described by  $\alpha$, $\beta$, $\xi$, we considered numerical  solutions of the initial-value problem for (\ref{master-radial0}) 
with initial data according to the asymptotic formula (\ref{master-solution_small_r})  for sufficiently small $r$. We used the  shooting method with the single shooting parameter $\Omega^2$   to  check condition (\ref{eigenvalue}) yielding the correct asymptotic behavior (\ref{Phi_at_infty}) at infinity. This procedure was carried out for a set of different parameters $(M\mu, Q)$.
\end{itemize}
In particular, we found that the static SS configurations considered above are linearly unstable at least in some region of parameters, which includes, for fixed $M\mu$, sufficiently small values of $Q$. This is enough to state that corresponding configurations are  unstable; at the same time, regions which are linearly stable with respect to radial perturbations may be unstable with respect to another types of perturbations.

\begin{figure}
    \centering
    \includegraphics[width=.49\textwidth]{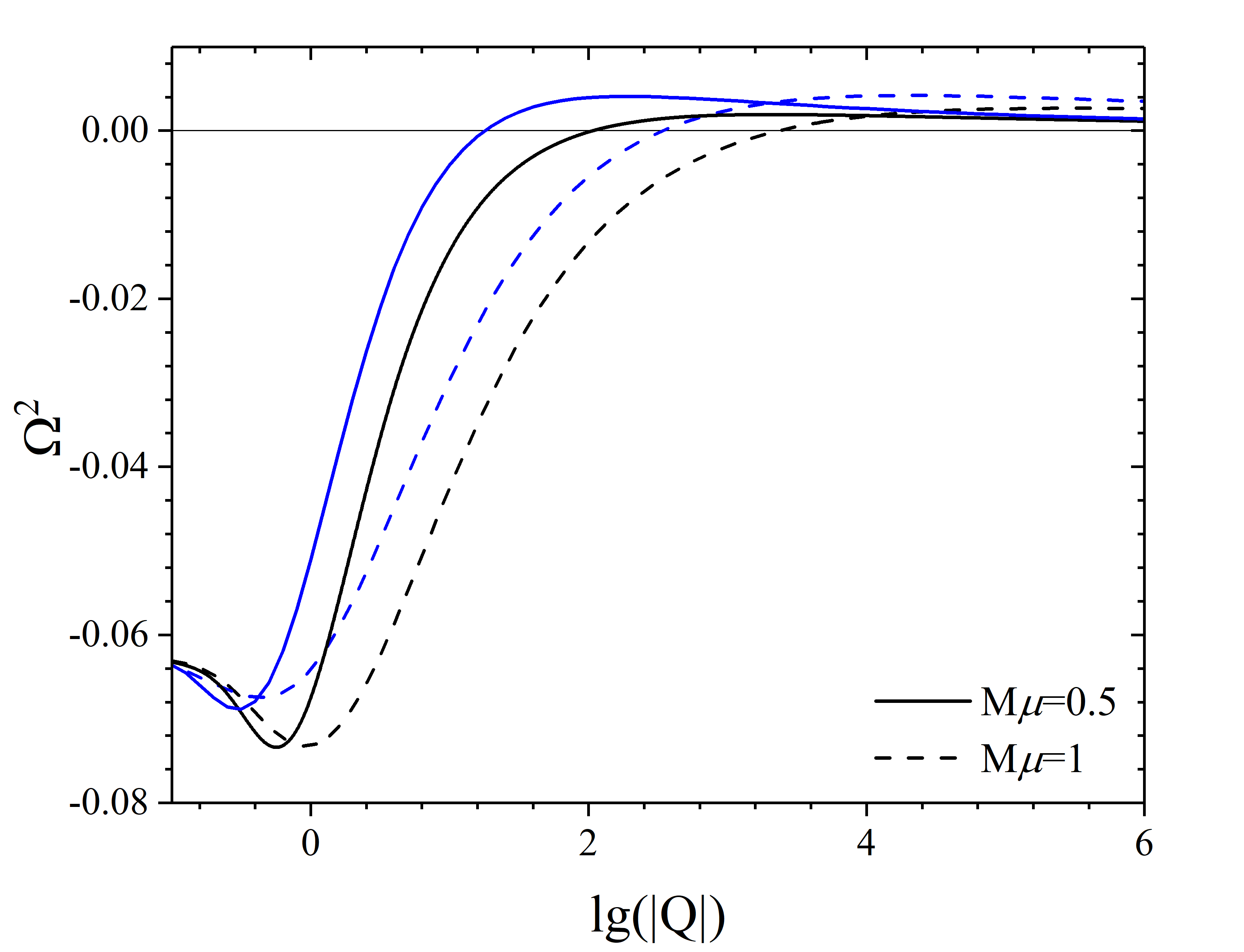}
    \caption{Typical dependencies of  $\Omega^2(Q)$. Black curves: $Q>0$; blue curves: $Q<0$.}
    \label{fig:Omega_of_Q}
\end{figure}

\begin{figure}
    \centering
    \includegraphics[width=.49\textwidth]{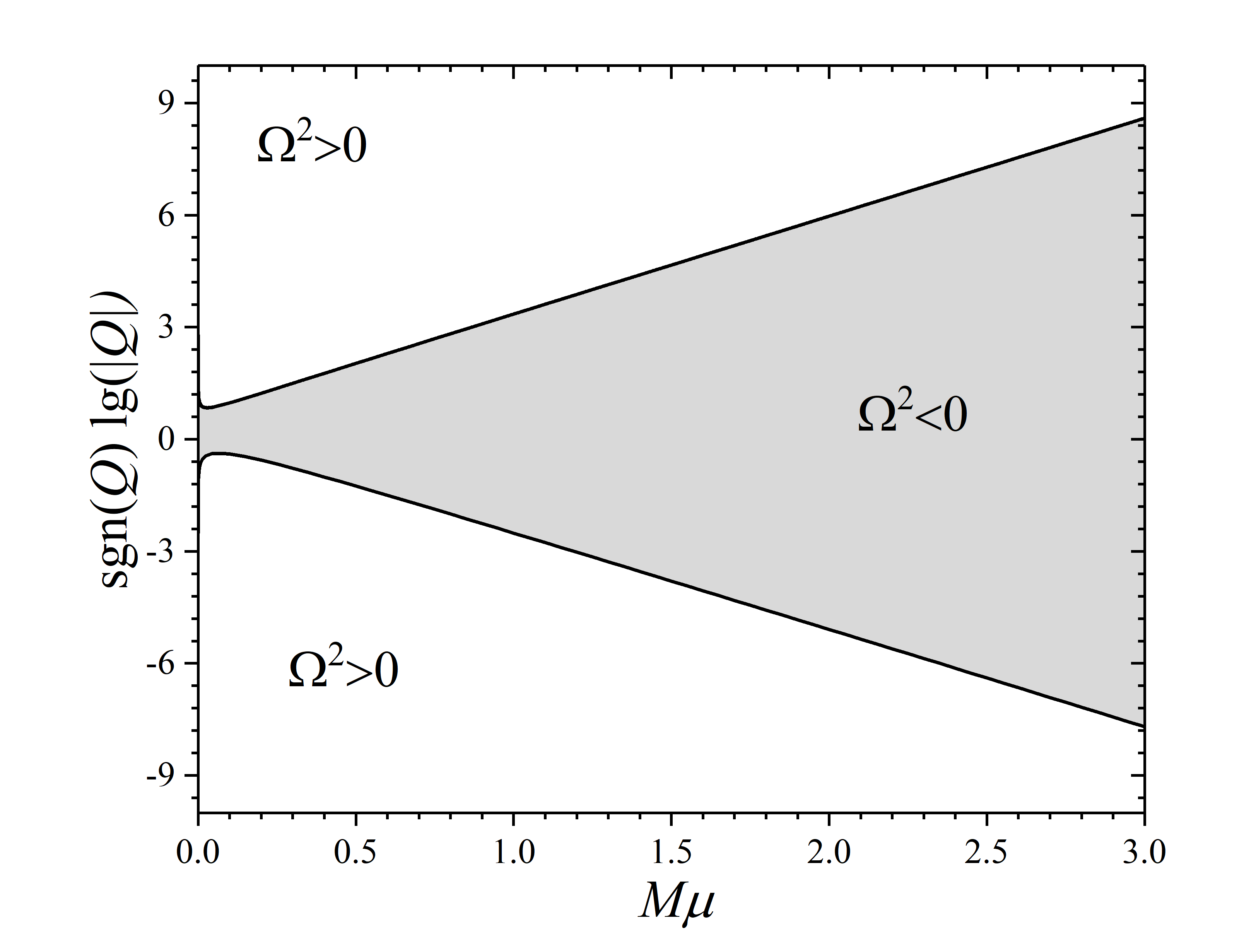}
    \caption{The grey region corresponds to linear instability against radial perturbations of the static SS configurations defined by the parameters $M\mu$ and $Q$. The boundary line corresponds to zero eigenvalues satisfying (\ref{eigenvalue}).}
    \label{fig:stabD}
\end{figure}

Fig.~\ref{fig:Omega_of_Q}  demonstrates  typical dependencies of the eigenvalues $\Omega$ as functions of $Q$.
The  domains of $(M\mu, Q)$,  which correspond  to linearly stable and unstable SS solutions against radial perturbations are  shown in Fig.~\ref{fig:stabD}.   
Transition to the Jordan frame preserves the time dependence $\sim e^{-i\Omega t}$ and, therefore, the stability/instability domains in  Fig.~\ref{fig:stabD}.


\section{Discussion}
\label{discussion}
 {We studied static asymptotically flat  spherically symmetric configurations   of the quadratic $f(R)$ gravity described by Eq.~(\ref{R2}). In this case, it is possible to introduce the scalaron field according to (\ref{conformal_xi}) and to apply the method of Section~\ref{sec:asympt}. In should be noted that this method uses   inequality (\ref{monotonocity}) that is rather restrictive;  it does not hold if, for example,  $f(R)$ contains terms  $\propto R^m$, $m>2$,  leading to decreasing  $W(\xi)$  for large $\xi>0$ (see, e.g., \cite{Shtanov-Sahni2023}).  Nevertheless, our method and results can be generalized to some class of $f(R)$ theories broader than (\ref{R2})  (see Appendix~\ref{generalizations} for examples).} 

Transition to the Einstein frame made possible purely analytical treatment of static global solutions using the method of \cite{ZhdSt}.
  {However, it is the metric $g_{\mu\nu}$ of the Jordan frame  that we consider to be the physical metric, responsible for all geometric measurements, whereas the conformally transformed  $\hat g_{\mu\nu}$ and the scalaron field are auxiliary objects. Indeed, we deal with different Riemannian spaces in Einstein and Jordan frames that may affect, e.g., the nature of singularities (see examples in \cite{Bahamonde-2016}). In this regard, we emphasize that the main  results of our  work concern the original Jordan metric.}

As follows from our findings, static asymptotically flat SS solutions with zero $T_{\mu\nu}$ are regular outside the center 
for any scalaron mass and for the global constants $\mathfrak{M}>0$,  $Q\ne 0$; they  do not have spherical singularities, which can arise, e.g., in   configurations with highly  nonlinear scalar-field potentials \cite{Strongly2021}.

In the case of an isolated  regular SS structure with a continuous mass-energy distribution in the central region, taking into account the experimental and theoretical bounds on the scalaron mass, the contribution of the scalaron field to the observable effects are expected to be very small. In this case, the gravitational field is determined by the energy-momentum tensor inside the body. Outside the SS body, the field is completely determined by two parameters: the mass $\mathfrak{M}$ and the ``scalar charge" $Q$. 

Returning to the case of  a purely gravitational configuration without ordinary matter ($T_{\mu\nu}\equiv 0$)  we state that the parameters $\mathfrak{M}$ and $Q$ act as the only characteristics of the static spherically symmetric system, and there is necessarily a naked singularity at the center  {if $Q\ne 0$}. In this regard,  {we confirm the statement  of  \cite{Hernandez2020} concerning the occurrence of NS at the center (although some provisions of this work need adjustment)}.  The only  exception is the Schwarzschild {black-hole solution that corresponds to zero scalaron field $\xi\equiv 0$ ($Q=0$); in this case, there is no difference between the Jordan and Einstein frames: $\hat g_{\mu\nu}\equiv g_{\mu\nu}$}. It is important to note that, whatever small nonzero  scalar charge could be, it strongly affects the  solution of  the quadratic $f(R)$ gravity \cite{Whitt:1984pd,Starobinsky:1980te}   in the interior region ($ r \lesssim r_g$) that is visible to an external observer and permeable to external signals moving towards the singularity. Apparently, this situation is typical for many isolated systems with a scalar field, ranging from the linear massless scalar field \cite{Fisher,JNW}   to  configurations with more complicated potentials \cite{ZhdSt}. No matter how small the scalar field is, it has a strong impact on the properties of spacetime in this region. At the same time, for large values of $Q$, the deviations from the Schwarzschild metric can be noticeable  at distances as large as, say, $r\sim 5$--$10\, r_g$. In any case, it is reasonable to defer detailed discussion of the corresponding  observational  effects until stability issues are resolved.

In this paper, we were not able to propose a  {completely} analytic treatment of stability, and we used numerical methods. 
We derived equations for linear perturbations and studied the asymptotic properties of their solutions near the center and for large $r$. As it is pointed out in Section \ref{SS_perturbations}, the physical conditions near the naked singularity are unknown; for example, the singularity may be radiating, absorbing or something else. However, we are limited to the assumption of small perturbations, which leads us to the  to the restriction (\ref{master-solution_small_r}) and  corresponds to the null Dirichlet condition at the center. With this condition we found exponentially growing modes of perturbations (satisfying boundary conditions at the center and at infinity) for some $M\mu, Q$ shown in the grey region of Fig. \ref{fig:stabD}. The static configurations with $\mu, Q$ from this region are unstable with respect to linear perturbations.  Although our numerical simulations involve moderate values of   $Q$ and $M\mu$, they suggest that the domain of instability becomes  larger for large  $M \mu$ so as to include astrophysically interesting values. 

On the other hand, there exists the region of parameters, where linear perturbations are bounded (the white region in Fig.~\ref{fig:stabD}). This  could  indicate the existence of stable configurations, if not for the following circumstances: (i)~in case of a nonlinear system, the existence of bounded linear perturbations does not always mean stability, and nonlinear corrections must be studied (cf., e.g., \cite{Gibbons2005}); (ii)~other types of linear perturbations (polar and axial ones) are also possible. 
In fact, there are general results on the linear stability against  axial perturbations leading to a Schrödinger-type equation   \cite{stashko2023quasinormal}. One can show that the wave potential in this equation appears to be positive-definite in case of our problem; this can be used  to prove stability with respect to axial linear perturbations. However,   investigation of  polar perturbations is mandatory. 
 
\vskip3mm \textit{Acknowledgements.}
O.S. is grateful to Igor Klebanov for kind hospitality at Princeton University. V.I.Z.  acknowledges partial support under the scientific program  ``Astronomy and space physics" of the Taras Shevchenko National University of Kyiv (project 0124U001660). The work of Yu.V.S. was supported by the Simons Foundation and by the National Academy of Sciences of Ukraine under project 0121U109612.
\\

\appendix
\section{Asymptotic properties as $r\to\infty$}\label{r_to_infty}

Here we prove the  {existence of the unique solutions} of (\ref{18-19})--(\ref{equation-xi-vacuum}) with conditions (\ref{asympt-r-to-infty}), (\ref{xi_inf}); at the same time, our aim is to  justify formula  (\ref{asy-phi-infty}). 
The proof uses the standard techniques  of the theory of differential equations used in considerations of the conditional stability \cite{Coddington1955}. We will isolate the linear part of the equations using specific properties of the system, separate the nonlinear terms  and reduce the problem to application of the contraction mapping principle.     The specific feature of our  treatment is that  we reduce the problem to integral equations so that to have expressions like (\ref{asy-phi-infty}) at the first iteration.

Further we use a new field variable $\psi=r\xi$; also it is natural to rewrite (\ref{18-19}), (\ref{18+19})  using the variables 
\begin{equation}
x=\frac{\alpha+\beta}{2}\,,\quad y=\frac{\beta-\alpha}{2}+\ln \left(1-r_g/r\right)\,.
\label{alpha-beta}
\end{equation}
This gives  
\begin{align}
    \frac{dx}{dr} &=F_2(r,\psi, d\psi/dr) \, ,
    \label{linear_part_x} \\
    \frac{dy}{dr}+\frac{x+y}{r-r_g} &=F_3(r,x,y,\psi, d\psi/dr)\,, 
    \label{linear_part_y}
\end{align}
where the nonlinear terms (starting from the second order in $x$, $y$, $\psi$, and $d\psi/dr$),  are collected into  $F_2$ and $F_3$.
Due to (\ref{asympt-r-to-infty}), we have 
\begin{equation} 
x(\infty)=y(\infty)=0\,. \label{x_y_at_infty}
\end{equation} 

Using (\ref{18+19}), we write Eq.~(\ref{equation-xi-vacuum}) for static SF in the form 
  \begin{eqnarray} 	 
 \frac{d^2\psi}{dr^2}  +\left(\frac{d\psi}{dr}-\frac{\psi}{r}\right)\left[\frac{1}{r}\left(1-e^\beta\right)+re^\beta W(\xi)\right] \nonumber \\ =\frac{r}{6}  e^{  \beta} W'(\xi)   \,.
    \label{wave_eq_psi}
 \end{eqnarray}
Here, in the first order, we have $e^\beta=1-r_g/r$ and $W'(\xi)=6\mu^2\xi$. 
After some rearrangement, we collect the linear part  of (\ref{wave_eq_psi}) into the left-hand side:
\begin{eqnarray}
 \frac{d^2\psi}{dr^2}  -\left(\frac{d\psi}{dr}-\frac{\psi}{r}\right)   \frac{r_g}{r(r-r_g)}- \frac{\mu^2\psi}{1-r_g/r} \nonumber \\ =   F_1(r,\alpha,\beta,\psi, d\psi/dr) \,,
\label{linear_part}	 
\end{eqnarray}  
where $F_1$ contains the nonlinear terms  in $x$, $y$, $\psi$, and $d\psi/dr$.

For arbitrary $n=1,2, \ldots$,  one can find an asymptotic solution\footnote{In fact, we look for an asymptotic solution of the linear homogeneous part of (\ref{linear_part}) defined by the left-hand side. The nonlinear corrections due to the right-hand side of (\ref{linear_part}), which are exponentially smaller than the terms of (\ref{asymptotic_solution}) as $r\to \infty$, are ``automatically" taken into account in the iteration process below.}   of (\ref{linear_part}) in the form
\begin{equation} 	 
v(r,\mu)=e^{-\mu r}\left(\frac{r_g }{r}\right)^m\left(1+\sum_{k=1}^n\frac{b_k(\mu)}{r^k}\right)\,,
\label{asymptotic_solution}	
\end{equation}
    $m\equiv m(\mu)=M\mu $, which satisfies the \textit{exact} relation 
\begin{equation} \label{asymptotic_expansion1}	  \frac{d^2v}{dr^2}  -\left(\frac{dv}{dr}-\frac{v}{r}\right)   \frac{r_g}{r(r-r_g)}   -\frac{\mu^2v}{1-r_g/r}   =   \frac{p_n(r,\mu)v}{r^{n+2}} \,,
\end{equation}
where   $p_n(r,\mu)$ is a rational function of $r$, bounded as $r\to\infty$;  the coefficients $b_k$ are determined recursively.  In particular, in the case of formula (\ref{asy-phi-infty}), we have  $n=2$; the validity of (\ref{asymptotic_expansion1}) can be checked by direct substitution.  

Denote 
\begin{equation}  
\psi_1(r)=v(r,\mu)\,,\quad \psi_2(r)=v(r,-\mu)\sim e^{\mu r} \left(\frac{r}{r_g}\right)^{m}\,.
\label{psi_2}
\end{equation}
The  equation for $\psi_2$ is obtained  from (\ref{asymptotic_expansion1}) by the change $\mu\to-\mu$ leading also to $m\to -m$. Now we have the exponentially decaying approximate solution  of (\ref{linear_part}) and exponentially growing  one. The   Wronskian of $\psi_1$, $\psi_2$ is calculated directly to be $w_{12}=2\mu \left[1+{\cal O} (r^{-1})\right]$.  

We use the method of variation of constants by setting 
\begin{subequations}
\begin{align}
\psi(r) &= K_1(r)\psi_1(r)+K_2(r)\psi_2(r)\,, \label{psi_by_A}\\
\frac{d\psi}{dr} &= K_1(r)\frac{d\psi_1}{dr}+ K_2(r) \frac{d\psi_2}{dr}\,, \label{Dpsi_by_A}
\end{align}
\label{variation}
\end{subequations}
where the ``variable constants" are 
$K_1$ and $K_2$. Unlike the usual method of variation of constants,  functions $\psi_1,\psi_2$ are not exact solutions of the linear part of (\ref{linear_part}), but satisfy (\ref{asymptotic_expansion1}) with different signs of $\mu$. 

Let $Q_*=Qr_g$. In order to have the asymptotics $\psi\sim Q_*\psi_1(r)$ as $r\to\infty$, corresponding to (\ref{asy-phi-infty}), we require 
\begin{equation}
K_1(\infty)=Q_*\,, \quad K_2(\infty)=0\,.
\label{K_asympt}
\end{equation}
A consequence of (\ref{variation}) is 
\begin{equation}
\frac{dK_1}{dr} \psi_1(r)+\frac{dK_2}{dr}\psi_2(r)  =0 \, .
\label{variation_DK}
\end{equation}
Now we differentiate   (\ref{Dpsi_by_A}) and  substitute the result   into (\ref{linear_part}) using (\ref{asymptotic_expansion1}) and  (\ref{variation}). We obtain 
\begin{equation}
\frac{dK_1}{dr} \frac{d\psi_1}{dr}  +\frac{dK_2}{dr} \frac{d\psi_2}{dr}  = -G(r,K_1,K_2)+F_1 \, ,
\label{variation_DDK}
\end{equation}
where 
\begin{equation}
G(r,K_1,K_2)= \frac{p_n(r,\mu)K_1\psi_1+p_n(r,-\mu)K_2\psi_2}{r^{n+2}} \,.
\label{denote2a}
\end{equation} 
Combination of (\ref{variation_DK}) and (\ref{variation_DDK})  yields separate 
   equations   for ${dK_1}/{dr}$ and  ${dK_1}/{dr}$ and then, by virtue of (\ref{K_asympt}), leads to integral equations
\begin{align} 
 K_1(r) &=Q_{*}-\int\limits_r^\infty  \frac{G(t)-F_1(t)}{w_{12}(t)} \psi_2(t)dt \,,
\label{A12a} \\    
 K_2(r) &=\int\limits_r^\infty  \frac{G(t)-F_1(t)}{w_{12}(t)} \psi_1(t)dt \,.
\label{A12b}
\end{align}

Equations (\ref{linear_part_x}), (\ref{linear_part_y}) with conditions (\ref{x_y_at_infty}) lead to the integral equations
\begin{align}
    x(r) &=-\int\limits_r^\infty F_2(t) dt\, ,
    \label{linear_part_x_int} \\ 
      y(r) &=-\int\limits_r^\infty \left[\frac{(t-r)F_2(t)+(t-r_g)F_3(t)}{r-r_g}  \right]dt \,. 
    \label{linear_part_y_int}
\end{align}
The integral equations (\ref{A12a}), (\ref{A12b}), (\ref{linear_part_x_int}), (\ref{linear_part_y_int}) can be presented in terms of new variables $u\equiv \{u_j\}$, $j=1,\ldots,4$,  {by means of the following substitutions}:
\begin{equation}
 K_1(r)=u_1(r)\,,\quad K_2(r)= u_2(r)\frac{\psi_1(r)}{\psi_2(r)} \, , 
  \label{u_1,u_2}
\end{equation}
and 
\begin{equation}
x(r) = u_3(r)\psi_1(r)\,, \quad  
y(r) =u_4(r)\psi_1(r)\,.  
\label{u_3,u_4}
\end{equation}
Accordingly, we have 
\begin{align*}
\psi(r) 
& =\left(u_1+u_2\right)\psi_1(r)\,,\\
\tilde G(r,u) &\equiv G(r,K_1,K_2)=\frac{p_n(r,\mu)u_1+p_n(r,-\mu)u_2}{r^{n+2}}\psi_1\,,
\end{align*}
 and the functions 
 \begin{equation*} 
\tilde F_j(r,u)\equiv F_j\left(r,x,y,\psi,\frac{d\psi}{dr}\right)\,, \quad j=1,2,3\, , 
\end{equation*}
should also be expressed in terms of  $u$ by means of (\ref{u_1,u_2})--(\ref{u_3,u_4}). 

Introduce the operator $\mathfrak{A}=\{\mathfrak{A}_1,\mathfrak{A}_2,\mathfrak{A}_3,\mathfrak{A}_4\}$, where
\begin{align}
 \mathfrak{A}_1(r,u) &=Q_*-\int\limits_r^\infty \frac{\tilde G(t,u)-\tilde F_1(t,u)}{w_{12}(t)} \psi_2(t)dt\,,
 \label{Int_eqs_1} \\[7pt]
\mathfrak{A}_2(r,u) &= \frac{\psi_2(r)}{\psi_1(r)}\int\limits_r^\infty \frac{\tilde G(t,u)-\tilde F_1(t,u)}{w_{12}(t)}  \psi_1(t)dt\,,
 \label{Int_eqs_2} \\
 \mathfrak{A}_3(r,u) &=-\frac{1}{\psi_1(r)}\int\limits_r^\infty \tilde F_2(r,u)dt\,,   \label{Int_eqs_3} \\
  \mathfrak{A}_4(r,u) &=-\int\limits_r^\infty  \frac{(t-r)\tilde F_2(t,u)+(t-r_g)\tilde F_3(t,u)}{\psi_1(r)(r-r_g)}   dt \,.
   \label{Int_eqs_4}
  \end{align}

With this definition, the operator equation  \begin{equation}
  u(r)=\mathfrak{A}(r,u)\,, 
   \label{Operator_eq}
  \end{equation}
 is equivalent to system (\ref{linear_part}), (\ref{linear_part_x}), (\ref{linear_part_y})  with 
conditions (\ref{xi_inf}) and (\ref{x_y_at_infty}) and, therefore,  to the system of equations (\ref{18-19})--(\ref{equation-xi-vacuum})    with conditions (\ref{asympt-r-to-infty}), (\ref{xi_inf}).
Equation (\ref{Operator_eq}) will be considered on interval  $[r_0,\infty)$, where 
   $r_0>r_g$ will be taken sufficiently large.  
 
Note that  
\begin{align} |\psi(r)|=|(u_1+u_2)\psi| &\le   |u(r)|_4 \psi_1(r)\,, \\
\left|\frac{d\psi}{dr}\right| &\le   \mu\left[1+{\cal O} (r^{-1})\right]|u(r)|_4 \,\psi_1(r)\,, \\
| \tilde G(r,u)| &\le \frac{C_1}{r^{n+2}}\psi_1(r)|u|_4\,, \quad C_1={\cal O} (1)\,,
\label{psi,by_u}
\end{align}
 $|u|_4= \sum\limits_{i=1}^4 |u_i| $, where   
we used the explicit expressions  (\ref{asymptotic_solution}), (\ref{psi_2}) and  representations (\ref{variation}).
Further, we use that, for sufficiently large $r$,
\begin{align*}
\psi_1(r)\psi_2(r) &=1+{\cal O} (r^{-1})\,,\nonumber\\ \int\limits_r^\infty \psi_1^2(t)dt &=\frac{\psi_1^2(r)}{2\mu}\left[1+{\cal O} (r^{-1})\right]\,.
\end{align*}

After  lengthy calculations, we obtain that the functions $\tilde F_j$ are Lipschitz continuous:
 \begin{equation} 
|\tilde F_j(r,u)-\tilde F_j(r,u')|\le H(u,u')|u-u'|_4\,,
\label{Lipshitz}
 \end{equation}
 where $j=1,2,3$,
  \begin{equation*} 
H(u,u')\le C_2\, \max\{|u|_4,|u'|_4\}\, \psi^2_1(r)\,,\quad C_2={\cal O} (1)\,.
 \end{equation*}
 Note that we have two small parameters: $\psi_1(r)$ and $(r_g/r)^{n+1}$; the contribution  of the first one to estimates (arising due to the  nonlinear terms $F_j$)  is exponentially smaller for  $r\ge r_0$. 
 Then, for sufficiently large $r_0$, we have  
 \[
\varepsilon(r_0)=\max\{(r_g/r_0)^{n+1},\psi_1(r_0)\}=  (r_g/r_0)^{n+1}\,.
  \]

Denote  $\|f\|_a^b\equiv \sup\{f(t), \, t\in[a,b]\}$.
Consider the Banach space $\mathcal{B}$ of bounded continuous functions on $[r_0, \infty)$ with the norm $\|f\|_{r_0}^\infty$ and a domain $\mathcal{D}\subset \mathcal{B}: \forall u \in \mathcal{D}$, $|u|\le 2|Q_*|$. The operator $\mathfrak{A}$ is well defined on  $\mathcal{D}$;  {all integrals are convergent}:  in (\ref{Int_eqs_1}),  { the integrand   is estimated as  $\sim r^{-{(n+2)}} $ due to (\ref{psi,by_u}) and (\ref{Lipshitz}); in  (\ref{Int_eqs_2})--(\ref{Int_eqs_4}), we have} exponentially decaying expressions in the integrands.
 
By virtue of the above estimates, we have, for  $u\in \mathcal{D}$,
\begin{align}
 |\mathfrak{A}(r,u)| \, &\le \, Q_*+C_3\varepsilon(r_0)\|u\|_{r}^\infty\,,
 \label{Int_eqs_itself} \\
 |\mathfrak{A}(r,u)-\mathfrak{A}(r,v)| \, &\le \, C_1 \varepsilon(r_0) {\|u-v\|_{r_0}^\infty} \,, 
 \label{Int_eqs_contract}
 \end{align}
where constants $C_1, C_3={\cal O} (1)$. Then from (\ref{Int_eqs_itself}) we have that (for sufficiently large $r_0$) $\mathfrak{A}(\mathcal{B})\subset \mathcal{B}$ and, from (\ref{Int_eqs_contract}), we infer the contraction property. Therefore, \textit{equation (\ref{Operator_eq}) has a unique solution}. 
 
This means  that there exists a unique solution $\alpha(r)\in C^1$, $\beta(r)\in C^1$, $\xi(r)\in C^2$   of  Eqs.~(\ref{18-19})--(\ref{equation-xi-vacuum})  with  conditions (\ref{asympt-r-to-infty}), (\ref{xi_inf}).

From (\ref{Int_eqs_1}), (\ref{Int_eqs_2})  we infer
\begin{align*}
|\psi(r)-Q_*\psi_1(r)|= |u_1(r)-Q_*+u_2(r)|\psi_1(r)  \\ \le \, {\cal O} \bigl( r^{-(n+1)} \bigr) \psi_1(r)\,.
\end{align*}
This confirms formula (\ref{asy-phi-infty}).  Also,  using (\ref{Int_eqs_3}), (\ref{Int_eqs_4}), we have
\begin{equation*}
|u_3(r)|+|u_4(r)|\lesssim \psi_1(r)\,,  
\end{equation*}
whence, on the account of (\ref{alpha-beta}),
\begin{align}
 \left| \alpha(r)-\ln\left(1-\frac{r_g}{r}\right) \right| + \left| \beta(r)+\ln\left(1-\frac{r_g}{r}\right) \right| \nonumber \\ \lesssim {\cal O} \left[ \psi^2_1(r) \right]\,.  
 \label{alpha-beta-first-iter}
 \end{align}
 Formula (\ref{asy-phi-infty}) is the result of the first iteration of (\ref{Operator_eq}). The next iterations add  higher-order corrections in powers of $1/r^{n+1}$ and $\exp\,(-\mu r)$, although the latter are useful only if very accurate results are required. In our study, the first iteration yielding (\ref{asy-phi-infty}) and (\ref{alpha-beta-first-iter}) was quite sufficient to impose the initial conditions for backward  numerical integration described in subsection~\ref{num_static}. 

\section{The case of $\xi<0$}\label{xi<0}

Here we use the method of \cite{ZhdSt}. The key question is to prove that  the right-hand sides of  (\ref{Ein_system-B}) and (\ref{Ein_system-C}) are bounded as $r\to 0$. 

Analogously to Section \ref{sec:asympt}, consider $r<r_1$ for some $r_1<\infty$.
If $\xi<0$ for $r\le r_1$, then    $0<W(\xi)<(3/4)\mu^2 \exp(4| {\xi}|)$, $|W'(\xi)|<3\mu^2 \exp(4 |\xi|)$, and $|\xi(r)|$ is monotonically decreasing due to (\ref{Ein_system-D}). For both right-hand sides (\ref{Ein_system-B}) and  (\ref{Ein_system-C}), it suffices  to estimate 
  \begin{equation}\label{P(r)}  
P(r)\equiv r^2 X(r)e^{4|\xi|} =r_1^2 X_1 e^{4|\xi_1|}e^{-2L+4d-(x_1-x) } \,,
 \end{equation}
  where $x\equiv x(r)\equiv \ln[X(r)]$, $r<r_1$,  and we denoted   $x_1=x(r_1)$,  $X_1=X(r_1)=\exp(x_1)$, $\xi_1=\xi(r_1)$, $d(r)=|\xi(r)|-|\xi_1|>0$, $L\equiv L(r)=-\ln(r/r_1)>0$.
  
We use the inequality 
 \[
\frac{[\xi(r)-\xi(r_1)]^2}{\ln\left( {r_1}/{r}\right)} \le  \int\limits_r^{r_1} t \left(\frac{d\xi}{dt}\right)^2dt\,,  \quad r<r_1\,,
 \]
 which is valid for any continuously differentiable function $\xi$ (see Eq.~(31) of \cite{ZhdSt}). By virtue of   (\ref{18-19}),    
 \begin{equation*}  
x_1-x=3\int\limits_r^{r_1}t\left(\frac{d\xi}{dt}\right)^2dt\ge  \frac{3}{L} d^2  \,.
 \end{equation*} 
    Then in (\ref{P(r)})
\begin{align*}
-2L+4d-(x_1-x) \le -2L+4d-\frac{3}{L}d^2 \\  
=-\frac{3}{L} \left( d-\frac23 L \right)^2-\frac23 L<0\,,  
\end{align*}
    whence   
 \begin{equation*}  
P(r)  \le  {r_1^2}X_1  (r/r_1)^{2/3}  e^{4|\xi_1|}\,.
 \end{equation*} 
 Therefore, the exponent $e^{4|\xi|}$ is suppressed by $X(r)$; the right hand sides of  (\ref{Ein_system-B}) and  (\ref{Ein_system-C}) are bounded.  
This is sufficient to infer the existence of the finite limits $Z_0$ and $Y_0$ for $r\to 0$. The considerations leading to the lower bounds $|Z_0|>0$ and $Y_0>0$ are similar to the corresponding arguments in Section \ref{sec:asympt} for positive $\xi(r)$.

\section{Asymptotic behavior near NS}\label{Near_NS}

Here we   formulate an iteration procedure to find the asymptotic expansion near NS in curvature coordinates. 
Taking into account (\ref{limit_z}) and (\ref{limit_Y}), we have from (\ref{Ein_system-B}), (\ref{Ein_system-C})
\begin{equation} \label{system-1Int} 
 Y(r)=Y_0+Y_1(r)\,, \quad  Z(r)=Z_0+Z_1(r)\,,    
\end{equation} 
where
\begin{equation}  \label{system-1Int-a} 
Y_1(r) =\int\limits_{0}^{r }X(t)[1- t^2W(\xi(t))]\frac{dt}{r_g}={\cal O} \left[ (r/r_g)^{\eta+1} \right],
\end{equation}
\begin{equation}
\label{system-1Int-b} 
   Z_1(r) = -  \frac{1}{6}\int\limits_{0}^{r } t^2 X(t)W'( \xi(t))  \frac{dt}{r_g}={\cal O} \left[ (r\mu)^2 (r/r_g)^{\eta+2\zeta+1} \right] .
\end{equation} 
Now we can separate the logarithmic terms in $x$, $\xi$ by setting
\begin{subequations}
\begin{align}
x(r)=  \eta\, \ln\left( \frac{r}{r_g}\right) +x^*_1(r)\, ,
\label{asympt_substitute_a} \\
\xi (r)= -\zeta\, \ln\left( \frac{r}{r_g}\right) +\xi^*_1(r)  \,, 
\label{asympt_substitute_b} 
\end{align}
\end{subequations} 
obtaining equations with smooth right-hand sides:
\begin{equation}  
\frac{d x^*_1}{d r}=\frac{3}{r}D_1(r)\,, \quad D_1(r)=\frac{Z^2}{Y^2}- \frac{Z_0^2}{Y_0^2}\,,
\label{chi-xi_1}
\end{equation}
\begin{equation}  
 \frac{d \xi^*_1}{d r}=\frac{D_2 (r)}{r}\,,\quad D_2(r) =\frac{Z_0}{Y_0}-\frac{Z}{Y}\,.      \label{chi-xi_2}
\end{equation}
On account of (\ref{system-1Int-a}), (\ref{system-1Int-b}), we have  $D_1(r)={\cal O} \left(r^{\eta+1}\right)$ and  $D_2(r)={\cal O} \left(r^{\eta+1}\right)$.

Then
\begin{equation} 
 x^*_1 (r)=\ln X_0 +3\int\limits_{0}^{r }\frac{dt}{t}D_1(t)\,,  
 \label{chi-xi_int_a}
 \end{equation}
 \begin{equation} 
 \xi^*_1  (r)=\xi_{0}+\int\limits_{0}^{r }\frac{dt}{t} D_2(t) \,, 
 \label{chi-xi_int_b} 
\end{equation}

 $X_0>0$ and $\xi_0$ being free constants.

Equations  (\ref{system-1Int-a}), (\ref{system-1Int-b}), (\ref{chi-xi_int_a}), (\ref{chi-xi_int_b}) with notation (\ref{system-1Int}), (\ref{asympt_substitute_a}),(\ref{asympt_substitute_b}) form  a system of integral equations for the    functions $Y_1$, $Z_1$, $\chi_1$, and $\xi_1$. This system is ready for successive approximations dealing with well-defined continuous expressions at each step of the iterative procedure that may be used to obtain the asymptotic solution of four equations (\ref{Ein_1-0})--(\ref{equation-phi}) with four constants  $Y_0$, $Z_0$, $X_0$, and $\xi_{0}$. 

The results of the first  iterations can be written as follows
\begin{equation}  \label{iterations-0} 
Y(r)=\ Y_0+ \frac{X_0 }{\eta+1}\left(\frac{r}{r_g}\right)^{\eta+1}\left[ 1+ {\cal O} \left(\mu^2r^2 \right)\right]\,, \\
\end{equation}
\begin{align} 
\label{iterations-1} 
x(r)&= \eta\, \ln \left( \frac{r}{r_g} \right) +\ln X_0 \nonumber\\
&{}-\dfrac{2\eta     }{(\eta+1)^2 }\dfrac{X_0}{Y_0}\left(\dfrac{r}{r_g}\right)^{\eta+1} \left[1+ {\cal O} \left(\mu^2r^2 \right)\right]\,, 
\end{align} 
\begin{align}
 \xi(r) &= -\zeta\, \ln \left( \frac{r}{r_g} \right) + \xi_0 \nonumber \\
 &{} + \dfrac{\zeta}{(\eta+1)^2}\dfrac{ X_0}{Y_0}\left(\dfrac{r}{r_g}\right)^{\eta+1}\left[1+ {\cal O} \left(\mu^2r^2 \right)\right] \,,     
\end{align}
where
  $\zeta=Z_0/Y_0$, $\eta =3 \zeta ^{2}$.

In case of a spatially flat configuration, the  constants  $Y_0$, $Z_0$, $X_0$, and $\xi_{0}$ are functions of the pair $M$, $Q$ that uniquely defines the configuration. Having the numerical solutions of Sec.~\ref{num}, we get $ Z_0$ of (\ref{limit_z}), $Y_0$ of (\ref{limit_Y}).  
We note a good numerical convergence to these limits.

Returning to the initial variables $e^\alpha$, $e^\beta$, we obtain the asymptotic formulas  (\ref{solution_as_ea})--(\ref{solution_as_xi}).


\section{Approximate SF for large $M\mu$ and small $\xi$}
\label{approximations-large_mu}
For $r>r_g$, as $r$ increases, we have $\xi\to 0$ and the metric reaches the Schwarzschild asymptotics. Moreover, as it was pointed out in Sec.~\ref{quadratic f(R) static}, the realistic values of $M \mu$ must be very large, so the  asymptotic behavior occurs very quickly.  Then we can use the WKB method in case of small $\xi\ll 1$, using (\ref{SF-potential_approx}) and   (\ref{equation-xi-small}). 
Substituting  $\xi= e^S$ into (\ref{equation-xi-small})  gives
\begin{equation}
\label{equation-S}
\frac{d^2S}{dr^2}+\left(\frac{dS}{dr }\right)^2+ \frac{dS}{dr }\frac{2r-r_g}{r(r-r_g)}   = \frac{r \mu^2}{r-r_g}    \,.   
\end{equation}
 Then we set
$S=S_0+S_1+ \ldots $, where the  first two orders  $|S_0|\sim {\cal O}(\mu)\gg|S_1|\sim {\cal O}(1)$ of the WKB method 
applied to (\ref{equation-S})  yield
\begin{equation}\label{phiWKB-N}
\xi(r)=  \frac{ {Q_1} \exp\left(-\mu r\sqrt{1-  {r_g}/{r} }\right)}{(1-{r_g}/{r})^{1/4}\left(1+\sqrt{1- {r_g}/{r}}\right)^{\mu r_g}}\left(  \frac{r_g}{r}\right)^{1+ M\mu}, 
\end{equation} 
 where we must put $Q_1=Q(4/e)^{ M\mu}$ so as to have correspondence with  (\ref{asy-phi-infty}) for $r\gg r_g$. 
Approximation (\ref{phiWKB-N}) is effective for $\xi(r)\ll 1$,  $\mu \sqrt{r(r-r_g)}\gg \ln(r/r_g)$. 
Note that formula (\ref{asy-phi-infty}),  unlike   (\ref{phiWKB-N}),  cannot be used for $r$ comparable with $r_g$. 

\section{Some generalizations}
\label{generalizations}
Our results on the global properties of static SS solutions can be generalized to a broader class of $f(R)$ gravity models if certain conditions are met. First of all, for unique determination of the dependence $u(\xi)$ we need  the monotonicity of $f'(R)$. Another  assumption is requirement (\ref{monotonocity}), which is needed to apply the method of Section~\ref{sec:asympt}. Without this asumption, a different  asymptotic behavior  at the center is possible.

From Eq.~(\ref{SF-potential}), we have 
 \begin{equation}
W'(\xi)=\frac{1}{\left[f'(u)\right]^2}\left[u f'(u)-2f(u)\right] \, , 
\label{DW}
\end{equation}
with the relation between $u$ and $\xi$ given by (\ref{SF-potential}). At the point of minimum (\ref{u_m}), this yields (\ref{mu^2}).

Taking into account that $\sgn\,\xi =\sgn \left(f'-1\right)$, we have that (\ref{monotonocity}) is satisfied if 
\begin{equation}
 \left(\frac{df}{du}-1\right) \left(u\frac{df}{du}-2f \right)>0\,.
 \label{monotonous-2}
\end{equation}
This inequality holds for the well-known models of \cite{Hu-Sawicky2007, Starobinsky-2_2007} at least in some ranges of parameter values.

Consider another example 
\begin{equation}
    f(u)=u-\frac{u^2}{6\mu^2}\frac{1+au^{2n}}{1+u^{2n}}, \quad 0<a<1\,,
    \label{example_f}
    \end{equation}
with the same scalaron mass $\mu$ as in the case of (\ref{R2}). 
    
In this case, there exists $u_0$ such that, for $u\in (-\infty,u_0)$, we have $df/du>0$ and $u(\xi)$ is uniquely defined; here, $u\to u_0$ corresponds to $\xi\to -\infty$, and $u\to -\infty$ to $\xi\to \infty$. For $n=1$, the requirements  $f''(u)\ne 0$  and (\ref{monotonous-2}) are fulfilled if $1/5< a\le1$; for $n=2$, the restriction is  $0.558< a\le 1 $. It is important to note that, in the case of (\ref{example_f}), we have an infinite ``shell''  $W(\xi)\to \text{const}$ providing suitable conditions for cosmological inflation \cite{Shtanov-Sahni2023}. As $\xi\to-\infty$, the behavior of solutions is also analogous to the case of (\ref{R2}).

\bibliography{references.bib}

\end{document}